\documentclass[twocolumn]{aastex631}  
\usepackage{amsmath}
\usepackage{graphicx}
\usepackage{amssymb}
\usepackage{natbib}
\usepackage{url}
\usepackage{color}
\usepackage{booktabs}
\usepackage{multirow}
\usepackage{xeCJK} 

\newcommand{\kms}{km\,s$^{-1}$}

\shorttitle{BHB Identification and Purification}
\shortauthors{Zhang, Shuai et al.}
\begin{document}
	
	\title{Photometric Identification and Dynamical Purification of Blue Horizontal Branch Stars in DESI MWS DR1}
	
	\author[0000-0003-2413-9587]{Shuai Zhang}
	\affiliation{School of Physics, Hebei Normal University, Shijiazhuang 050024, People's Republic of China}
	\affiliation{Guo Shoujing Institute for Astronomy, Hebei Normal University, Shijiazhuang 050024, People's Republic of China}
	
	\author[0000-0003-2536-3142]{Xiaolong Wang}
	\affiliation{School of Physics, Hebei Normal University, Shijiazhuang 050024, People's Republic of China}
	\affiliation{Guo Shoujing Institute for Astronomy, Hebei Normal University, Shijiazhuang 050024, People's Republic of China} 
	\affiliation{Physics Postdoctoral Research Station at Hebei Normal University}
	
	\author[0000-0001-8424-1079]{Kai Xiao}
	\affiliation{School of Astronomy and Space Science, University of Chinese Academy of Sciences, Beijing 100049, People's Republic of China}
	
	\author[0000-0003-1828-5318]{Guozhen Hu}
	\affiliation{School of Physics, Hebei Normal University, Shijiazhuang 050024, People's Republic of China}
	\affiliation{Guo Shoujing Institute for Astronomy, Hebei Normal University, Shijiazhuang 050024, People's Republic of China}
	\affiliation{Physics Postdoctoral Research Station at Hebei Normal University}
	
	\author[0000-0003-1359-9908]{Wenyuan Cui}
	\affiliation{School of Physics, Hebei Normal University, Shijiazhuang 050024, People's Republic of China}
	\affiliation{Guo Shoujing Institute for Astronomy, Hebei Normal University, Shijiazhuang 050024, People's Republic of China}
	
	\author[0000-0002-6540-2372]{Ce Cai}
	\affiliation{School of Physics, Hebei Normal University, Shijiazhuang 050024, People's Republic of China}
	\affiliation{Guo Shoujing Institute for Astronomy, Hebei Normal University, Shijiazhuang 050024, People's Republic of China}
	
	\author[0000-0002-9767-9237]{Shuai Feng}
	\affiliation{School of Physics, Hebei Normal University, Shijiazhuang 050024, People's Republic of China}
	\affiliation{Guo Shoujing Institute for Astronomy, Hebei Normal University, Shijiazhuang 050024, People's Republic of China}
	
	\author{Jie Ju}
	\affiliation{School of Sciences, Hebei University of Science and Technology, Shijiazhuang 050018, People’s Republic of China}
	
	\correspondingauthor{Wenyuan Cui}
	\email{cuiwenyuan@hebtu.edu.cn}
		
	\begin{abstract}
		Blue horizontal-branch (BHB) stars are ideal tracers for mapping the structure of the Galactic stellar halo. 
		However, traditional spectroscopic identification is limited by data quality and sample size, while broad-band photometric methods often achieve only moderate completeness and purity.
		Building upon the color-space BHB identification method of \citet{2025ApJS..278...62H}, we synthesize SkyMapper $(u,v,g,z)$ photometry from DESI MWS DR1 spectroscopy and identify 9,237 BHB candidates in the $(g-z)_0$--$(u-v)_0$ color space using the HDBSCAN clustering algorithm. 
		We then construct a Kernel Density Estimation (KDE) Bayesian probability model with capture-recapture (C-R) prior to assign posterior membership probabilities, achieving a probability-weighted purity of 88.9\% and completeness of 94.2\%. 
		Applying an escape-velocity cut based on the Galactic gravitational potential, we remove 1,461 contaminants with unphysical kinematic parameters, yielding a clean photometric BHB sample $\mathcal{S}_c$ of 7,776 stars (retention rate $R_{\rm keep} = 84.2\%$). 
		Systematic analysis of the contaminants reveals that they are predominantly faint sources ($g\sim20$~mag) with very low signal-to-noise ratios (median SNR~$<$~5). When the BHB color--luminosity relation is incorrectly applied to these objects, their space velocities are amplified beyond the Galactic escape limit, confirming that the escape-velocity filter essentially acts as a physical detector of low-SNR--driven distance inflation.					
		
	\end{abstract}
		

	\section{Introduction}\label{sec:intro}
	
	Blue horizontal-branch (BHB) stars are helium-core-burning stars located on the blue side of the RR Lyrae instability strip on the Hertzsprung–Russell diagram, with masses of about $0.5$--$1.0\,M_\odot$, effective temperatures $T_{\rm eff}\sim 7000$--$12,000$ K, strong Balmer jumps, and narrow yet prominent Balmer line profiles \citep{2009Ap&SS.320..261C, 2021ApJ...923..162R, 2024ApJS..270...11J, 2010ApJ...708L.121D, 2002MNRAS.337...87C, 2012AJ....143...86V}. 
	The $g$-band absolute magnitude of BHB stars is nearly constant ($M_g\sim0.5$--$0.7$) over a limited color range ($-0.3<g-r<0.0$), earning them the reputation of ``standard candles'' with a photometric distance precision of $\sim0.2$~mag (5\% distance uncertainty) \citep{1974ApJS...28..157G, 2004AJ....127..899S}. 
	As old ($\gtrsim12$~Gyr in globular clusters) and metal-poor tracers, BHB stars are widely present in globular clusters and the Galactic stellar halo \citep{1952AJ.....57....4A, 1955ApJS....2....1H, 2001PASP..113.1162M, 2015ApJ...813L..16S, 2010ApJ...708..698D}. 
	Compared with other halo tracers, BHB stars offer significant advantages: they are brighter and provide higher distance precision than main-sequence turn-off (MSTO) stars ($M_g\sim+0.7$ vs.\ $M_g\sim3.5$, distance precision $\sim5\%$ vs.\ $\sim9\%$), and they are more abundant than RR Lyrae variables \citep{2008ApJ...684.1143X, 2008ApJ...680..295B, 1998MNRAS.297..732S, 2011MNRAS.416.2903D}. 
	These properties make BHB stars ideal probes for mapping the three-dimensional structure of the Galactic stellar halo and unveiling its merger history.
	
	However, effectively distinguishing BHB stars from their primary contaminants — blue stragglers (BS), main-sequence A-type and very late main-sequence B-type stars, which have similar effective temperatures but higher surface gravities — has always been a core challenge. 
	Traditional spectroscopic methods separate the two populations through Balmer line profile analysis:  \citet{2002MNRAS.337...87C} proposed the $D_{0.15}$--color method and the scale-width--shape method, which can reliably separate BHB stars from BS stars when ${\rm H}\gamma>13$ \AA\ and the continuum S/N$\sim15$\,\AA$^{-1}$, with the equivalent width of the Ca II K line serving as an additional discriminant. 
	Although spectroscopic methods offer high accuracy, they are limited by observational efficiency; spectroscopically selected samples are typically small (a few thousand stars), making it difficult to conduct uniform all-sky surveys over large areas \citep{2008ApJ...684.1143X, 2011ApJ...738...79X, 2012AJ....143...86V}.
	
	Complementary to spectroscopic methods, photometric approaches exploit the sensitivity of blue filters to the Balmer jump to distinguish BHB stars from BS stars. 
	Early methods include the $(U-B)_0$--$(B-V)_0$ two-color diagram \citep{1991ApJ...375..121P}, the SDSS $u-g$--$g-r$ color cut that identified 1,493 BHB stars \citep{2000ApJ...540..825Y}, and the work of \citet{2012AJ....143...86V}, who utilized the sensitivity of Paschen features in the $griz$ bands to surface gravity to select BHB stars in surveys lacking a $u$ band. 
	However, the discriminatory power of broad-band photometry alone is limited: completeness is typically only 50\%--70\%, and the contamination rate can reach $\sim30\%$ \citep{2010AJ....140.1850B, 2018PASJ...70...69F, 2018MNRAS.481.5223T}. 
	Recently, \citet{2025ApJS..278...62H} used Gaia XP spectra to synthesize $(u-v)_0$ and $(g-i)_0$ colors in the SkyMapper photometric system and constructed a catalog of 49,733 BHB stars (hereafter Hu2025 BHBs) with both completeness and purity exceeding 90\%, covering nearly the entire sky. 
	These advances fully demonstrate the great potential of medium-band photometric and synthetic photometric methods in identifying large samples of BHB stars.
		
	In terms of kinematics, BHB stars provide key constraints for understanding the dynamical state of the Galactic halo. 
	\citet{1984ApJ...281..260P} found that halo BHB stars do not participate in the disk's rotation (with evidence of a slight retrograde motion), their velocity ellipsoid is moderately anisotropic (radially elongated), and the metal-poor sub-sample ([Fe/H]$<-2.0$) exhibits a higher solar motion. 
	\citet{2008ApJ...684.1143X} found that the line-of-sight velocity distribution of BHB stars within 10--60 kpc can be described by a Gaussian with $\sigma_{\rm los}\approx111$ km s$^{-1}$. 
	\citet{2011ApJ...738...79X} identified position-velocity substructures of high statistical significance in over 4,000 BHB stars, quantitatively confirming the expectations of hierarchical formation theory. 
	\citet{1999AJ....117.2329W} confirmed that field horizontal-branch stars and main-sequence A-type stars differ significantly in both metallicity and velocity distributions, and that the degree of cross-contamination is limited.
		
	Over the course of surveys, BHB star samples have undergone a dramatic expansion from the earliest efforts to the present day. 
	The earliest systematic studies of BHB stars can be traced back to \citet{1984ApJ...281..260P}, who performed a kinematic analysis of the halo using 150 southern AB stars; \citet{1991ApJ...380..104N}, who placed initial constraints on the dark corona mass using 9 distant BHB stars (40--65 kpc); and \citet{1999AJ....117.2329W}, who carried out a systematic spectroscopic classification of 444 field horizontal-branch stars. 
	Entering the SDSS era, \citet{2004AJ....127..899S} isolated 1170 BHB stars (contamination $<10\%$ for bright stars $g<18$, $\sim25\%$ for faint stars), \citet{2008ApJ...684.1143X} strictly selected 2401 BHB stars from SDSS DR6, and \citet{2011ApJ...738...79X} obtained 4985 BHB stars from SDSS DR8. 
	The SEGUE survey specifically designated BHB stars as a spectroscopic target class, providing four-dimensional information ($\alpha,\delta,d,V_{\rm los}$) with a distance precision of about 5\% and radial velocity precision of 5--20 km s$^{-1}$ \citep{2009AJ....137.4377Y,2011ApJ...738...79X}. 
	The LAMOST survey identified about 5000 BHB stars from its DR5 \citep{2024ApJS..270...11J}. 
	The DESI Milky Way Survey (MWS) can measure radial velocities for millions of stars with a precision of $\sim1$ km s$^{-1}$ and provide metallicity measurements with an accuracy of $\sim0.2$ dex, wherein BHB stars are prioritized for observation owing to their ability to trace distant regions \citep{2023ApJ...947...37C}; \citet{2025MNRAS.542..560B} constructed the largest spectroscopic BHB sample to date based on DESI data — 9866 stars extending to 120 kpc, with the number of stars at Galactocentric distances $>50$ kpc being four times that of previous catalogs. 
	\citet{2008AJ....135..564B} used the Century Survey to identify 655 BHB stars from 2414 2MASS candidates, covering 4300 square degrees.
	
	In this context, the present work is motivated by the convergence of two recent developments. First, DESI MWS DR1 provides moderate-resolution spectroscopy for millions of stars with precise radial velocities ($\sim1$ km s$^{-1}$) and metallicities ($\sim0.2$ dex), covering a large fraction of the sky, yet its spectroscopically confirmed BHB sample remains limited to a few thousand objects \citep{2023ApJ...947...37C, 2025MNRAS.542..560B}. Second, the synthetic photometry method of \citet{2025ApJS..278...62H} demonstrated that medium-band colors, specifically $(u-v)_0$ and $(g-i)_0$ synthesized from Gaia XP spectra, can separate BHB stars from blue stragglers and main-sequence A stars with both completeness and purity exceeding 90\%. In this work, we find that replacing $(g-i)_0$ with $(g-z)_0$ further improves the separation in the DESI MWS DR1 sample, because the $z$-band contains surface-gravity-sensitive Paschen features that enhance the contrast between BHB stars and high-gravity contaminants (see Figure~\ref{fig:color_color}). By synthesizing SkyMapper photometry directly from DESI MWS DR1 spectra, we can apply this improved color-space selection to a much larger sample than the spectroscopically confirmed one, while retaining DESI's homogeneous spectral information for kinematic and chemical follow-up. This combination naturally leads to the present approach: we use HDBSCAN density-based clustering in the $(g-z)_0$--$(u-v)_0$ color space to identify BHB candidates, and then apply a Bayesian probabilistic model and a dynamical escape-velocity filter to purify the sample.
	
	Looking ahead, near-ultraviolet photometry from Chinese Space Station Telescope (CSST) is expected to extend BHB star detection to the outer halo and even the edge of the Galaxy \citep{2025ApJS..278...62H}; combined with data from spectroscopic surveys such as DESI and LAMOST and with Gaia astrometry, this will provide an unprecedented data foundation for studies of the structure and evolution of the Milky Way \citep{2023ApJ...947...37C}.
	
	This paper is organized as follows. 
	Section~2 describes the DESI MWS DR1 data, the synthesis of SkyMapper photometry from the spectra, and the HDBSCAN clustering method used to identify BHB candidates in the $(g-z)_0$--$(u-v)_0$ color space. 
	Section~3 introduces a kernel density estimation Bayesian probability model, combined with a  C-R  framework, to assign posterior membership probabilities to each candidate and evaluate the purity and completeness of the photometric BHB sample.
	Section~4 presents the dynamical purification based on an escape-velocity cut derived from the Galactic gravitational potential, yielding a clean photometric BHB sample $\mathcal{S}_c$.
	Section~5 investigates the statistical properties of the contaminants removed by this cut.
	Section~6 provides an independent purity validation using SDSS globular cluster H-R diagrams.
	Finally, Section~7 summarizes the main findings and discusses future prospects.

	\section{BHB Identification} \label{sec:BHBIdentification}
	\subsection{DATA}
	
	DESI MWS is an essential component of the DESI survey and represents one of the largest and most efficient multi-object spectroscopic surveys currently conducted on 4-meter-class telescopes. Its advantages include extensive sky coverage, large target numbers, low selection bias, moderate spectral resolution, and full wavelength coverage, enabling precise measurements of radial velocity ($\approx 1$ km/s) and metallicity ([Fe/H] $\approx$ 0.2 dex) \citep{2023ApJ...947...37C}. Compared with previous surveys like SEGUE and LAMOST, DESI MWS has made significant improvements in faint star sample density, outer halo detection capability, and chemical abundance measurements. It is particularly suited for studying the Milky Way's thick disk, stellar halo, dark matter distribution, and the history of galaxy accretion.
	
	Our analysis is based on the spectroscopic data from the Milky Way Survey (MWS) within the Data Release 1 (DR1) dataset of the DESI survey \citep{2023ApJ...947...37C}. Via the merged spectrum from the B, R, and Z bands ($\lambda \approx 3600–9800\,\text{\AA}$), we obtain extinction-corrected spectrum by applying the extinction parameter $E(B–V)$ from the all-sky reddening map SFD98 \citep{1998ApJ...500..525S} and the F99 extinction model \citep{1999PASP..111...63F} . Subsequently, using the filter response functions of the SkyMapper photometric system, we convolve the corrected spectrum with specific filters ($u, v, g, r, i, z$) to compute the corresponding AB magnitudes in each band. This method ultimately provides extinction-corrected six-band model magnitudes for each spectrum, which can be used for subsequent stellar classification and astrophysical parameter studies.
	
	Particularly, we must retain the $u$ band (by trimming the SkyMapper $u$ filter to only the part with $\lambda > 3600\,\text{\AA}$, which is the lower limit of the complete DESI spectra) because this band is crucial for the identification of blue horizontal-branch (BHB) stars and is one of the key features to distinguish BHB stars from other blue stellar populations \citep{2008ApJ...684.1143X}.
	
	This work employs a combination of $(u-v)_0$ and $(g-z)_0$ color parameters to identify and classify Blue Horizontal Branch (BHB) stars. The theoretical basis stems from the differential sensitivity of different photometric bands to stellar atmospheric parameters: effective temperature ($T_{\mathrm{eff}}$), surface gravity ($\log g$), and metallicity ([Fe/H]).
	
	The $u$ band covers the Balmer jump region ($\lambda = 3646$\,\AA) in its response function. The amplitude of this jump is highly sensitive to stellar surface gravity ($\log g$): For giants with lower $\log g$ (e.g., Blue Horizontal Branch stars, BHB, $\log g \sim 2.8$--3.75), their relatively thin atmospheres result in weaker pressure broadening effects, enhancing the prominence of the Balmer jump and reducing the $u$-band flux. In contrast, for main-sequence stars with higher $\log g$ (e.g., Blue Stragglers, BS, $\log g \sim 3.75$--5.0), pressure broadening smooths out the jump, making the $u$ band appear brighter \citep{2002MNRAS.337...87C, 2008AJ....135..564B}. Additionally, the $u$ band contains numerous metal absorption lines, so its flux is also influenced by metallicity. Consequently, the color index $(u-v)_0$, after partially compensating for metallicity effects, can serve as an effective tracer of $\log g$: lower $\log g$ corresponds to larger (redder) $(u-v)_0$ values.
	
\begin{figure*}[htbp]
		\centering
		\includegraphics[width=0.8\linewidth]{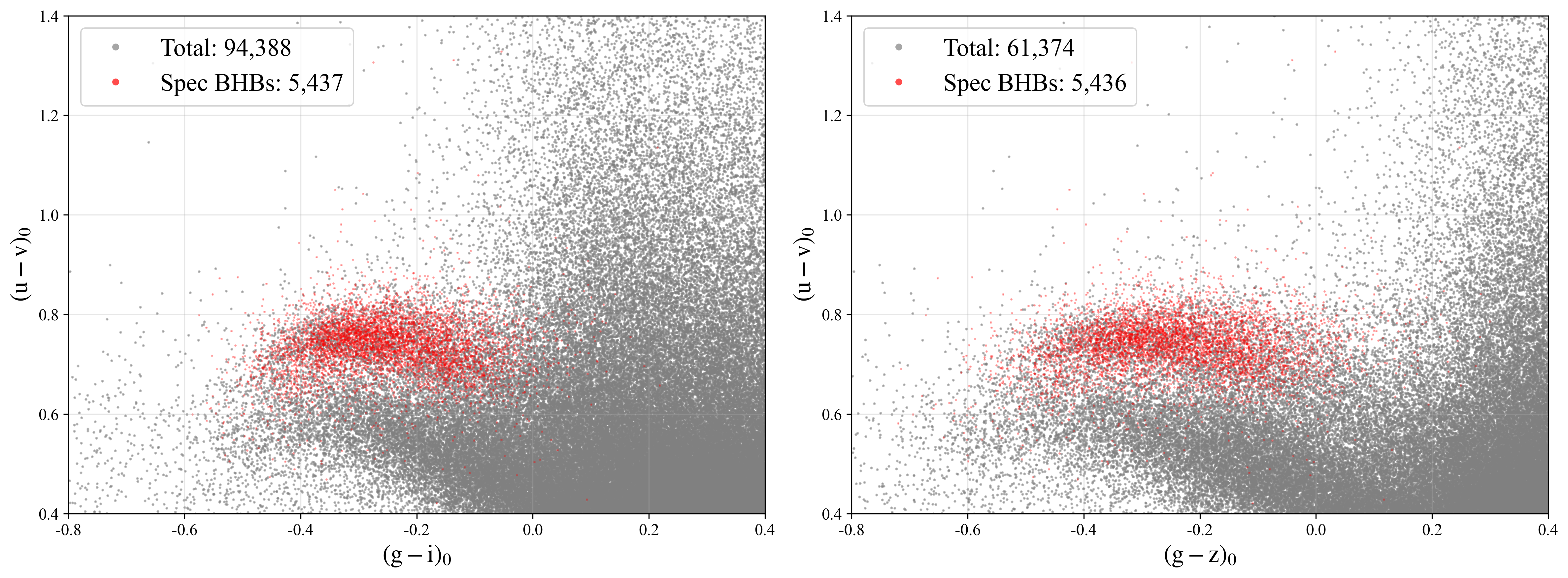}
		\caption{
			Distribution of BHB stars in color space. The left panel shows the distribution in the $(g-i)_0$ vs. $(u-v)_0$ color space, and the right panel shows the distribution in the $(g-z)_0$ vs. $(u-v)_0$ color space. Red dots represent the DESI DR1 spectroscopically confirmed BHB sample from \citet{2025MNRAS.542..560B}. The BHB cluster is more clearly separated from the non-BHB distribution in $(g-z)_0$.
		}
		\label{fig:color_color}
	\end{figure*}

	Furthermore, we adopt $(g-z)_0$ rather than the commonly used $(g-i)_0$ as an indicator in that the Paschen features in the $z$-band are sensitive to surface gravity \citep{2012AJ....143...86V, 1998ApJS..119..121L}. In the two-dimensional color space of $(u-v)_0$ versus $(g-z)_0$, BHB stars (low $\log g$, moderate $T_{\mathrm{eff}}$) and other stars (high $\log g$, lower $T_{\mathrm{eff}}$) will be distributed in distinct regions. Due to their low $\log g$, BHB stars occupy a region with larger $(u-v)_0$, while others with high $\log g$, are located in a region with smaller $(u-v)_0$. Meanwhile, the difference in $T_{\mathrm{eff}}$ further separates the two populations along the $(g-z)_0$ axis. Just as shown in left and right panels of Figure \ref{fig:color_color}, color-color diagrams constructed based on the DESI MWS DR1 spectroscopic survey \citep{2023ApJ...947...37C}, the gray points represent the color distribution of the total MWS data sample, while the red points denote the BHB sample spectroscopically confirmed in the DESI DR1 dataset \citep{2025MNRAS.542..560B}.  It is evident that the distribution of BHB stars is quite concentrated, and more separated from other stars along the $(g-z)_0$ axis than  $(g-i)_0$. Therefore, we filter the sample by applying the selection criteria  as below:
	\begin{enumerate}
		\item[(a)] $0.4 <(u-v)_0<1.4$ and $-0.8<(g-z)_0<0.4$
		\item[(b)] $\lvert b \rvert \geq$ 15$^{\circ}$ to avoid disk stars.
	\end{enumerate}
	This approach allows us to retain the majority of BHB candidates while significantly reducing the sample size. Except for a few outliers, this region essentially encompasses the vast majority of the DESI DR1 spectroscopically confirmed BHB sample \citep{2025MNRAS.542..560B}. Ultimately, a total sample $\mathcal{N}$ ($|\mathcal{N}| = 61{,}374$) remains, and all subsequent analyses are based on this sample.
	
	\subsection{Identification of BHB Stars}
	
	\begin{figure}[htbp]
		\centering
		\includegraphics[width=0.8\linewidth]{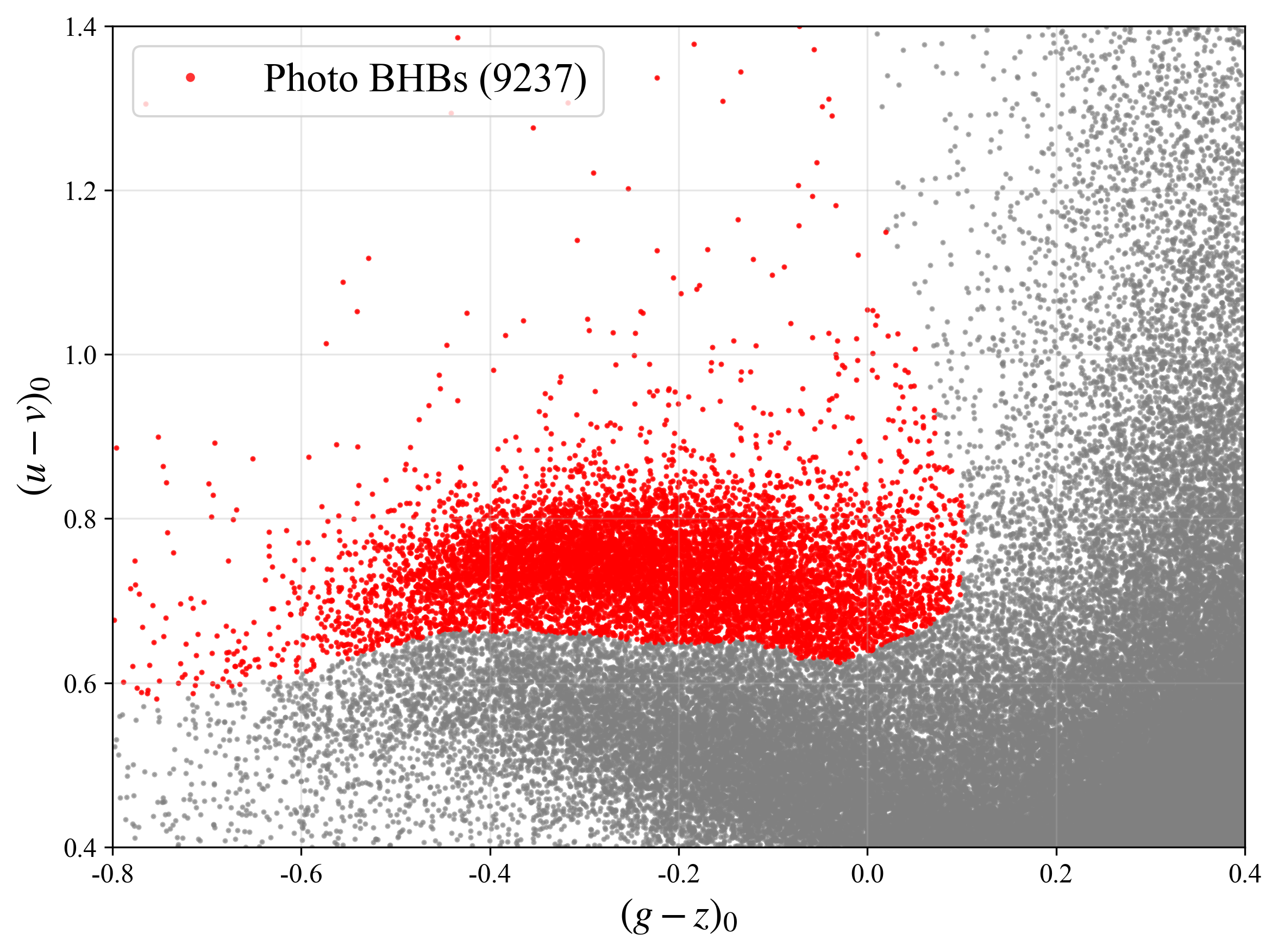}
		\caption{
			Using the HDBSCAN clustering algorithm to classify the sample, the identified BHB stars (red points) are clearly concentrated in a tight cluster, while the gray points have a significantly different density distribution from the BHB stars.
		}
		\label{fig:BHB_cluster}
	\end{figure}
		
	As shown in the right panel of Figure \ref{fig:color_color}, the BHB stars form a compact clump embedded in a more diffuse non-BHB population. To identify this overdensity quantitatively, we apply the HDBSCAN clustering algorithm \citep{2017JOSS....2..205M} in the two-dimensional color space spanned by $(u-v)_0$ and $(g-z)_0$.
	Prior to clustering, the two color indices are standardized to zero mean and unit variance, so that the Euclidean metric treats both axes equally despite their different intrinsic dispersions.
	HDBSCAN constructs a density-based hierarchy using the mutual reachability distance, with $\texttt{min\_samples}=20$ serving as a local density smoothing scale. The parameter values are chosen according to the apparent structure of the color--color diagram: $\texttt{min\_cluster\_size}=3000$ is set to be sufficiently large that only the prominent BHB overdensity is retained as a cluster, while smaller-scale density fluctuations are suppressed; $\texttt{min\_samples}=20$ provides a local density smoothing scale that is insensitive to the sparse outskirts of the BHB clump; $\texttt{cluster\_selection\_epsilon}=0.05$ (in standardized color units) is deliberately kept small so that the BHB clump is not merged with the surrounding diffuse non-BHB population.
	Because the BHB clump is the most prominent density peak in this color space, with a membership ($\sim 10^4$) far exceeding the adopted $\texttt{min\_cluster\_size}$, and because it is separated from the surrounding non-BHB population by a clear density gap, the resulting cluster is inherently robust to moderate variations of these parameters.
	Because the BHB clump sits on a shallow density gradient, genuine BHB stars near its periphery may initially be labeled as noise. We therefore reassign each noise point to its nearest cluster: for each noise point, the five nearest clustered members are identified, exponential weights $\exp(-d^2/0.5)$ are assigned according to their distances, and the point is assigned to the cluster with the highest total weight.
	The resulting cluster , as indicated by the red points in Figure \ref{fig:BHB_cluster}, is coincides with the compact BHB overdensity visible in the color--color diagram: it contains 9237 stars, of which 4968 belong to the DESI DR1 spectroscopically confirmed BHB sample \citep{2025MNRAS.542..560B}; we denote this cluster as the photometric BHB candidate set $\mathcal{C}$.
		
	For the convenience of the subsequent analysis and discussion, we adopt the following set definitions:
	\begin{enumerate}
		\item[$\mathcal{K}$:] the DESI DR1 spectroscopically confirmed BHB sample from \citet{2025MNRAS.542..560B}, $|\mathcal{K}| = 5436$; 5 additional spectroscopically confirmed BHB stars are excluded because they deviate too far from the BHB cluster region in the color diagram.
		
		\item[$\mathcal{C}$:] the full set of stars identified as the BHB cluster in the $(g-z)_0$--$(u-v)_0$ space, with a sample size of $|\mathcal{C}| = 9237$.
	\end{enumerate}
		
	\section{Evaluation of the Clustering Result}\label{sec:prob_model_eval}
	\subsection{Data Preprocessing}
	
	Based on the total sample $\mathcal{N}$, which includes the photometric color-parameter clustered BHB sample $\mathcal{C}$ and the spectroscopically confirmed BHB sample $\mathcal{K}$, in order to conduct a systematic analysis specifically for the BHB sample identified by photometric clustering, we define the following sets:
	\begin{enumerate}
		\item[$\mathcal{S}_1$:] BHB stars that are both spectroscopically confirmed and photometrically identified, with $|\mathcal{S}_1| = 4968$;
		\item[$\mathcal{S}_2$:] BHB stars that are only spectroscopically confirmed ($|\mathcal{S}_2| = |\mathcal{K}| - |\mathcal{S}_1| = 468$);
		\item[$\mathcal{S}_3$:] BHB stars that are only photometrically identified ($|\mathcal{S}_3| = |\mathcal{C}| - |\mathcal{S}_1| = 4269$);
		\item[$\mathcal{S}_4$:] Stars classified as non-BHB by both methods ($|\mathcal{S}_4| = |\mathcal{N}| - |\mathcal{K}| - |\mathcal{C}| + |\mathcal{S}_1| = 51{,}669$).
	\end{enumerate}
	
	We apply the C-R method using the Lincoln-Petersen estimator, which gives
	\begin{eqnarray}
		T_{\mathrm{LP}} &=& \frac{|\mathcal{K}| \times |\mathcal{C}|}{|\mathcal{S}_1|} \sim 10{,}107\\
	\end{eqnarray}

	Although the C-R method assumes independence between the two captures, in our case the two methods are highly correlated, implying that the C-R estimate $\hat{T}$ serves only as a conservative lower bound on the true total number of BHBs. Nevertheless, $\hat{\pi} = \hat{T}/|\mathcal{N}| = 0.1647$ still serves as a good conservative prior estimate, upon which we construct a probabilistic model.
	
	\subsection{Constructing a Probabilistic Model to Evaluate Clustering Results}\label{sec:Probabilistic_Model}
	
	To evaluate the performance of the clustering algorithm in identifying BHB stars, we construct a probabilistic model based on Kernel Density Estimation (KDE) to quantify the posterior probability of each sample being a BHB star. The core idea is to use the positive and negative sample sets to establish probability density models in the feature space, and then compute the posterior probability $P(\text{BHB}|\mathbf{x})$ via Bayes' theorem. The specific steps are as follows:
	\begin{enumerate}	
		\item \textbf{Training preparation}: We use the four mutually exclusive subsets defined in the C-R  estimation. Specifically, $\mathcal{S}_1$ serves as the positive sample set, containing $|\mathcal{S}_1| = 4968$ stars, and $\mathcal{S}_4$ serves as the negative sample set, containing $|\mathcal{S}_4| = 51{,}669$ stars. $\mathcal{S}_2$ and $\mathcal{S}_3$ are not used for model training and are reserved solely for subsequent evaluation.
		
		\item \textbf{Kernel density estimation}: We use a Gaussian kernel to build probability density models for the positive and negative samples separately:
		\begin{eqnarray}
			\hat{p}(\mathbf{x}|\text{BHB}) &=& \frac{1}{n_{\text{pos}} h_{\text{pos}}^d} \sum_{i=1}^{n_{\text{pos}}} K\left(\frac{\mathbf{x} - \mathbf{x}_i^{\text{(pos)}}}{h_{\text{pos}}}\right) \label{eq:bhb} \\
			\hat{p}(\mathbf{x}|\text{nonBHB}) &=& \frac{1}{n_{\text{neg}} h_{\text{neg}}^d} \sum_{i=1}^{n_{\text{neg}}} K\left(\frac{\mathbf{x} - \mathbf{x}_i^{\text{(neg)}}}{h_{\text{neg}}}\right) \label{eq:nonbhb} 
		\end{eqnarray}
		
		Bandwidth selection follows the Silverman rule:
		\begin{equation}
			h = \left( \frac{4}{d+2} \right)^{\frac{1}{d+4}} \cdot n^{-\frac{1}{d+4}} \cdot \hat{\sigma}
		\end{equation}
		
		where the Gaussian kernel is defined as:
		\begin{equation}
			K(\mathbf{u}) = \frac{1}{(2\pi)^{d/2}} \exp\left(-\frac{1}{2}\mathbf{u}^T\mathbf{u}\right)
		\end{equation}
		
		Here, $n_{\text{pos}} = |\mathcal{S}_1| = 4968$ and $n_{\text{neg}} = |\mathcal{S}_4| = 51{,}669$ are the numbers of positive and negative samples, respectively, and $d = 2$ is the data dimension. $\hat{\sigma} = \frac{1}{d} \sum_{j=1}^{d} \sigma_j$ is the average of the standard deviations across dimensions. The calculated bandwidths are $h_{\text{pos}}=0.0223$ for the positive sample and $h_{\text{neg}}=0.0328$ for the negative sample.
		
		\item \textbf{Posterior probability calculation}: Based on the conservative prior probability $\hat{\pi} = \hat{T}/|\mathcal{N}| = 0.1647$ obtained from the C-R  estimation, we apply Bayes' theorem to compute the posterior probability for each star:
		\begin{eqnarray}
			P(\text{BHB}|\mathbf{x}) &=& \frac{\hat{\pi} \cdot \hat{p}(\mathbf{x}|\text{BHB})}{\hat{\pi} \cdot \hat{p}(\mathbf{x}|\text{BHB}) + (1-\hat{\pi}) \cdot \hat{p}(\mathbf{x}|\text{nonBHB})} \label{eq:pos_bhb} \\
			P(\text{nonBHB}|\mathbf{x}) &=& 1 - P(\text{BHB}|\mathbf{x}) \nonumber \\ 
			&=& \frac{(1-\hat{\pi}) \cdot \hat{p}(\mathbf{x}|\text{nonBHB})}{\hat{\pi} \cdot \hat{p}(\mathbf{x}|\text{BHB}) + (1-\hat{\pi}) \cdot \hat{p}(\mathbf{x}|\text{nonBHB})} \label{eq:pos_nonbhb}
		\end{eqnarray}
		
		\item \textbf{Calibration}: The initial model exhibits overconfidence, with the posterior probability distribution tending toward extreme values (close to 0 or 1). We apply temperature scaling \citep{2017arXiv170604599G} to calibrate the model:
		\begin{eqnarray}
			P_{\text{c}}(\text{BHB}|\mathbf{x}_i) &=& \frac{P(\text{BHB}|\mathbf{x}_i)^{1/T}}{P(\text{BHB}|\mathbf{x}_i)^{1/T} + (1-P(\text{BHB}|\mathbf{x}_i))^{1/T}} \label{eq:c_bhb}\\
			P_{\text{c}}(\text{nonBHB}|\mathbf{x}_i) &=& 1 - P_{\text{c}}(\text{BHB}|\mathbf{x}_i) \\\nonumber
			&=&  \frac{(1-P(\text{BHB}|\mathbf{x}_i))^{1/T}}{P(\text{BHB}|\mathbf{x}_i)^{1/T} + (1-P(\text{BHB}|\mathbf{x}_i))^{1/T}} \label{eq:c_nonbhb}
		\end{eqnarray}
		
		Because the calibration set is highly imbalanced, selecting $T$ by directly minimizing the Expected Calibration Error (ECE)  is unreliable: the global average is dominated by the numerous low-probability negative samples, which conceal overconfidence near the decision boundary.  Yet overconfidence in the raw posteriors is clearly present (the distribution is sharply peaked at 0 and 1).  We therefore adopt $T=1.5$, under which the global ECE remains very low (0.011), while the probability distribution is moderately broadened to alleviate boundary overconfidence.  The model still retains high  probability-weighted purity and completeness.
		
		\item \textbf{Purity and Completeness}: Based on the calibrated posterior probabilities, we compute the probability-weighted purity and completeness. Here, the photometric cluster $\mathcal{C}$ plays the role of the positive class in a hard classification, while each member $\mathbf{x}_i \in \mathcal{C}$ contributes to the true-positive count with a weight equal to its calibrated posterior probability $P_{\rm c}(\text{BHB}|\mathbf{x}_i)$. The probability-weighted purity is therefore the expected fraction of true BHB stars within $\mathcal{C}$, i.e., the probability-weighted analogue of the standard binary classification purity, $TP/(TP+FP)$, where $TP$ and $FP$ denote the numbers of true and false positives, respectively. Likewise, the probability-weighted completeness is the expected fraction of all true BHB stars in the full sample $\mathcal{N}$ that are contained in $\mathcal{C}$, i.e., the probability-weighted analogue of the standard completeness (or recall), $TP/(TP+FN)$, where $FN$ denotes the number of false negatives. In the limit where all posterior probabilities are either 0 or 1, these probability-weighted measures reduce to the conventional hard-classification definitions:
		\begin{align}
			P_{\text{prob}} &= \frac{\sum_{i \in \mathcal{C}} P_{\text{c}}(\text{BHB}|\mathbf{x}_i)}{|\mathcal{C}|} \label{eq:purity}\\
			C_{\text{prob}} &= \frac{\sum_{i \in \mathcal{C}} P_{\text{c}}(\text{BHB}|\mathbf{x}_i)}{\sum_{i=1}^{|\mathcal{N}|} P_{\text{c}}(\text{BHB}|\mathbf{x}_i)} \label{eq:completeness}
		\end{align}

		where $|\mathcal{C}| = |\mathcal{S}_1| + |\mathcal{S}_3| = 9237$ is the total number of photometric BHB stars, and $|\mathcal{N}| = 61{,}374$ is the total sample size.
		
	\end{enumerate}
	
	\begin{figure*}[htbp]
		\centering
		\includegraphics[width=\textwidth]{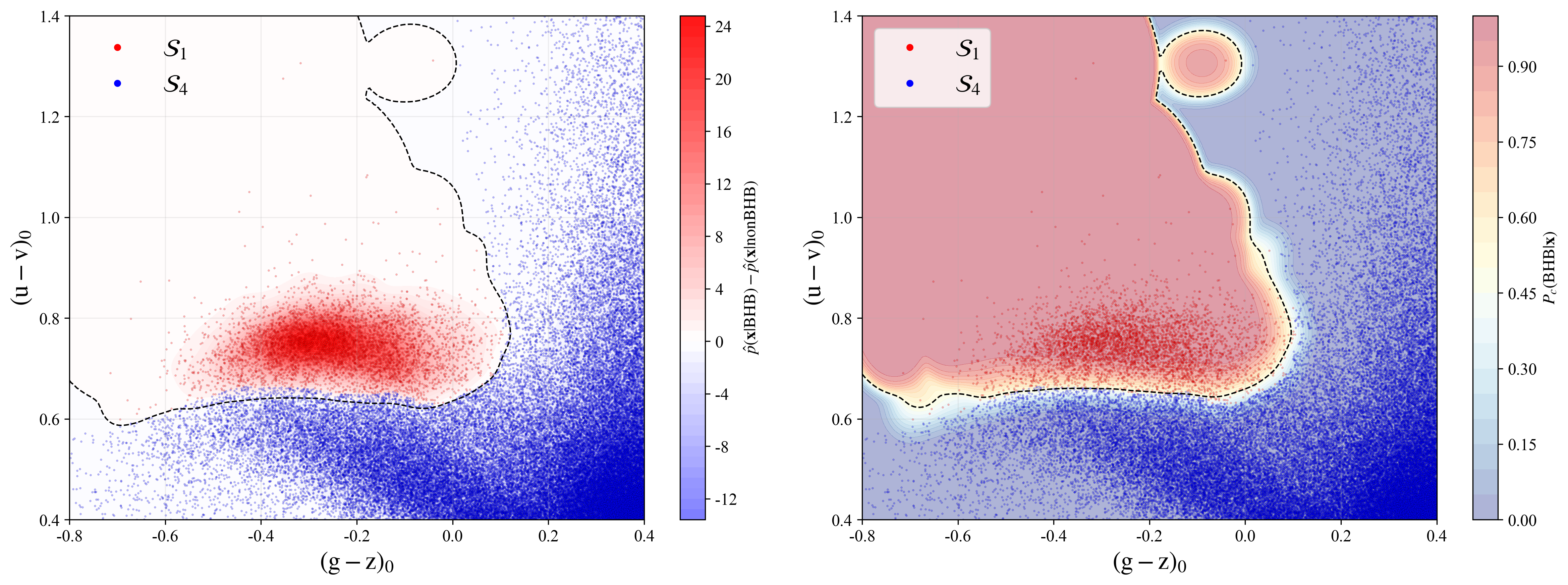}
		
		\caption{\textbf{Left panel:} Likelihood density difference, i.e., the difference between the kernel density estimates based on the positive BHB sample ($\mathcal{S}_1$) and the negative non-BHB sample ($\mathcal{S}_4$): $\hat{p}(\mathbf{x}|\text{BHB}) - \hat{p}(\mathbf{x}|\text{nonBHB})$. Red regions indicate higher density for positive samples (more likely to be BHB), blue regions indicate higher density for negative samples (more likely to be non-BHB), and the black dashed line marks the zero contour. This panel does not incorporate the prior probability, showing only the likelihood comparison between the two classes in the feature space ($(g-z)_0$, $(u-v)_0$). \textbf{Right panel:} Calibrated posterior probability distribution  $P_{\text{c}}(\text{BHB}|\mathbf{x}_i)$, i.e., the probability of each feature point being a BHB star. Colors ranging from blue to red represent probabilities from 0 to 1, and the black dashed line indicates the 0.5 contour (classification boundary). This panel shows the spatial distribution of the model's confidence in BHB membership.}
		\label{fig:probability_difference}
	\end{figure*}
	
	The left panel of Fig.~\ref{fig:probability_difference} shows the difference between the kernel density estimates of the BHB and nonBHB samples in the feature space, i.e., the subtraction of Eq.~\ref{eq:bhb} and Eq.~\ref{eq:nonbhb}:
	\begin{equation}
		\Delta(\mathbf{x}) = \hat{p}(\mathbf{x}|\text{BHB}) - \hat{p}(\mathbf{x}|\text{nonBHB})
		\label{eq:probability_difference}
	\end{equation}
	where the positive region (red) indicates a higher relative density for BHB, and the negative region (blue) indicates a higher relative density for nonBHB. This density difference map differs from the posterior probability map in that it contains no prior information; it merely reflects the relative likelihood density distribution of the positive and negative samples ($\mathcal{S}_1$ and $\mathcal{S}_4$) in the feature space. The zero-level contour (black dashed line) represents the boundary where the two class densities are equal. This boundary does not perfectly coincide with the separation between the positive and negative sample data. The primary reason is that near the concave side of the bow-shaped positive sample distribution, the boundary lies close to the core of the positive sample region; due to KDE smoothing, the likelihood probability density decays more slowly in the concave direction than the actual data density, pushing the zero-level contour into the negative sample region. Additionally, differences in bandwidth and the incompleteness of the training data (only $\mathcal{S}_1$ and $\mathcal{S}_4$ included) also lead to slight discrepancies between the two.
	
	By combining the likelihood density distributions with the prior probabilities estimated from C-R analysis, we obtain the posterior probabilities $P(\text{BHB}|\mathbf{x})$ (Eq.~\ref{eq:pos_bhb}) and $P(\text{nonBHB}|\mathbf{x})$ (Eq.~\ref{eq:pos_nonbhb}). However, the posterior probabilities computed directly from Bayes' rule tend to be overconfident. Temperature scaling can calibrate and flatten this overly sharp distribution, leading to more reasonable confidence levels for samples near the classification boundary in the parameter space. The calibrated posterior probability distribution (Eq.~\ref{eq:c_bhb}) is shown in the right panel of Fig.~\ref{fig:probability_difference}. In this panel, the red region indicates a high probability of BHB, the blue region corresponds to a low BHB probability (i.e., high nonBHB probability), and the black dashed line marks the decision boundary, which largely coincides with the boundary of the BHB sample cluster identified in the clustering analysis. The BHB and nonBHB sample points are overlaid on the plot; their distribution in color space shows strong agreement with the BHB discrimination provided by the calibrated probability model, indicating that the calibrated posterior probability model is highly reliable.
	
	\begin{figure*}[htbp]
		\centering
		\includegraphics[width=0.8\textwidth]{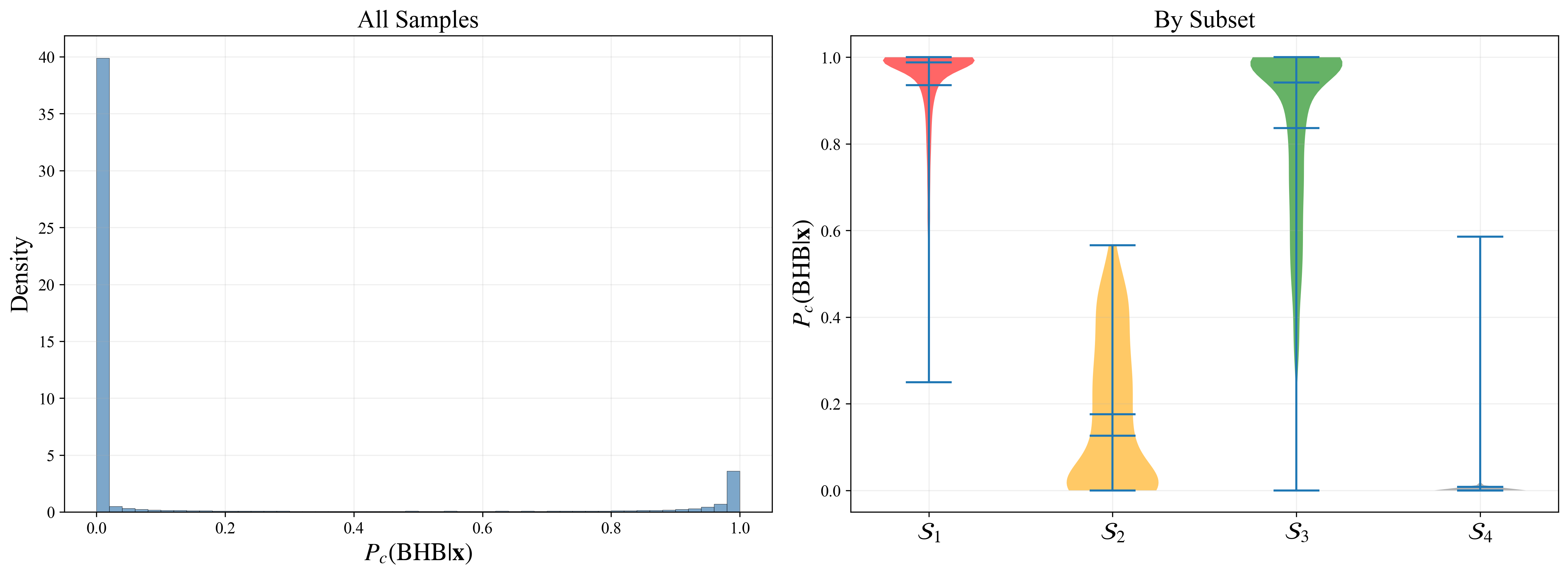}
		\caption{Left panel: Histogram of the calibrated posterior probability density $P_{\text{c}}(\text{BHB}|\mathbf{x})$ for all stellar samples. The distribution exhibits a bimodal structure, with the majority of samples concentrated in the low-probability region ($P_{\text{c}}(\text{BHB}|\mathbf{x}) \approx 0$), reflecting the predominance of nonBHB stars in the data. Approximately 8\% of the samples have high posterior probabilities ($P_{\text{c}}(\text{BHB}|\mathbf{x}) \approx 1$), corresponding to the BHB candidates identified by the clustering method. Right panel: Violin plots of the calibrated posterior probability distribution for the four mutually exclusive subsets. Both $\mathcal{S}_1$ and $\mathcal{S}_3$ show very high posterior probabilities (median $>0.95$ for $\mathcal{S}_1$, median $>0.85$ for $\mathcal{S}_3$), whereas the distribution for $\mathcal{S}_4$ is concentrated in the extremely low probability range (median nearly 0). The distribution of $\mathcal{S}_2$ is scattered across relatively low probabilities, reflecting the probability uncertainty caused by inconsistent labels. The vast majority of samples in $\mathcal{S}_3$ have high posterior probability values, indicating that this subset as a whole is a reliable set of BHB candidates.}
		\label{fig:subsets_post_dist}
	\end{figure*}
	
	We then use the calibrated probability model to assess the confidence level of each sample in the classification outcome, i.e., by computing the probability that each sample is a BHB star and comparing the probability distributions across the different sample subsets to evaluate the classification performance.
	
	Figure~\ref{fig:subsets_post_dist} illustrates the distributional characteristics of the calibrated posterior probability $P_{\text{c}}(\text{BHB}|\mathbf{x})$. The left panel shows a histogram of the posterior probability density for all samples, which exhibits a typical bimodal distribution: one peak near $P_{\text{c}}(\text{BHB}|\mathbf{x}) \approx 0$ (nonBHB samples) and the other near $P_{\text{c}}(\text{BHB}|\mathbf{x}) \approx 1$ (BHB samples). This clear separation indicates that the probabilistic model constructed from the selected feature parameters (the $(g-z)_0$ and $(u-v)_0$ color indices) can effectively distinguish BHB stars from nonBHB stars. The right panel displays violin plots of the posterior probability distributions for the four mutually exclusive subsets:
	\begin{itemize}
		\item $\mathcal{S}_1$: The median posterior probability is as high as 0.9881, indicating that these BHB samples—jointly confirmed spectroscopically and identified by photometric clustering in this work—have extremely high confidence.
		\item $\mathcal{S}_2$: The median posterior probability is only 0.1266. These are BHB samples that are spectroscopically confirmed but were not recognized by photometric clustering; they account for approximately 8.6\% of the total official spectroscopic BHB sample ($468/5436$). In the color parameter space, these sources overlap with other stellar types and cannot be separated by clustering, and are thus assigned low confidence.
		\item $\mathcal{S}_3$: The median posterior probability is 0.9416, suggesting that the new BHB candidates discovered by the clustering algorithm also possess relatively high confidence.
		\item $\mathcal{S}_4$: The median posterior probability is close to 0, indicating high purity of the negative sample set. This is partly because we have already excluded BHB candidates as thoroughly as possible, and partly because even if a small number of BHBs remain, they are insufficient to affect the overall probability distribution given the large size of this dataset.
	\end{itemize}
	
	The hierarchical structure of the posterior probability distributions across the subsets confirms the validity of the clustering results: the clustering method not only successfully identifies the majority of known spectroscopic BHBs, but also uncovers a substantial number of high-confidence candidate samples. Meanwhile, the spectroscopic BHB samples missed by the clustering algorithm (accounting for 8.6\% of the known BHBs) generally exhibit low posterior probabilities, suggesting that these sources may indeed be difficult to distinguish from background stars in the feature space.
	
	Based on the current model (Eq.~\ref{eq:purity}--\ref{eq:completeness}) and the available data, we find that the purity of the photometric BHB sample from the clustering results is $88.99\%$, and the completeness is $94.2\%$. Furthermore, unlike conventional binary hard classification, which assigns each object a discrete label (BHB or non-BHB) without an associated uncertainty, the calibrated posterior probability model provides a soft classification: each source is assigned a continuous probability $P_{\rm c}(\text{BHB}|\mathbf{x}_i) \in [0,1]$ that quantifies the confidence of class membership. In the limiting case where all probabilities are rounded to 0 or 1, this soft classification reduces to the usual hard classification. The probability values can be used directly to select high-confidence samples according to a chosen threshold (e.g., by applying a high $P_{\rm c}$ threshold) or as weights in subsequent statistical analyses, thereby preserving information about classification uncertainty that would be lost under a hard-label assignment.

\section{Dynamical Purification Based on the Galactic Gravitational Model}\label{sec:de_contamination}

We compute the distance of BHBs using the color--distance correlation 	from \citet{2025ApJS..278...62H}, based on the $g$-band absolute magnitude 	calibration $M_g = -1.43\,(g-i)_0 + 0.23\,[\mathrm{Fe/H}] + 0.49$, where $(g-i)_0$ is the extinction-corrected colour and $[\mathrm{Fe/H}]$ is the DESI MWS metallicity. The extinction corrections use $A_g = 3.54\,E(B-V)$ and $A_i = 2.48\,E(B-V)$ with $E(B-V)$ from SFD98 \citep{1998ApJ...500..525S}. The heliocentric distance follows from the distance modulus: $d = 10^{(g_0 - M_g + 5)/5}\,\mathrm{pc}$.

Using the celestial coordinates, distance, radial velocity, and proper motions of the sample, we compute Galactocentric cylindrical coordinates $(R, z, \phi)$ and cylindrical velocity components: $V_R$ (radial velocity), $V_\phi$ (azimuthal velocity), and $V_z$ (vertical velocity). The total space velocity of each star is synthesized from the three-dimensional cylindrical velocity components: $v_{\rm total} = \sqrt{V_R^2 + V_\phi^2 + V_z^2}$.

We then employ the Galactic gravitational potential model of \citet{2017MNRAS.465...76M} to compute the local escape velocity $v_{\rm esc}(R, z)$ at the Galactocentric cylindrical coordinates $(R, z)$ for all BHB sample stars. At the solar position ($R_\odot = 8.21$~kpc, $z \approx 0$), this potential model yields an escape velocity of about 403~km/s. If a BHB candidate satisfies $v_{\rm total} > v_{\rm esc}(R, z)$, it cannot be gravitationally bound by the Milky Way, implying that the distance and kinematic parameters derived from the BHB standard candle are inaccurate, and this BHB candidate can thus be identified as a contaminant.

\begin{table}[htbp]
	\centering
	\caption{Statistics of Unbound Stars ($v_{\rm total} > v_{\rm esc}$)}
	\label{tab:contamination}
	\begin{tabular}{lrrr}
		\hline
		\textbf{Sample} & \textbf{Total} & \textbf{Unbound}  & \textbf{Rate} \\
		\hline
		$\mathcal{C}$ & 9237 & 1461 & 15.8\% \\
		$\mathcal{K}$ & 5436 & 364 & 6.7\% \\
		\hline
		$\mathcal{S}_1$ & 4968 & 283 & 5.7\% \\
		$\mathcal{S}_2$ & 468 & 81 & 17.3\% \\
		$\mathcal{S}_3$ & 4269 & 1178 & 27.6\% \\
		\hline
	\end{tabular}
\end{table}

Table~\ref{tab:contamination} lists the numbers and fractions of unbound stars in each sample and subsample. 
The contamination rates exhibit a markedly asymmetric distribution among the subsamples:
$\mathcal{S}_1$ (5.7\%) and $\mathcal{S}_2$ (17.3\%) have relatively low contamination rates, whereas $\mathcal{S}_3$ reaches a high fraction of 27.6\%, contributing 80.6\% of the unbound stars in $\mathcal{C}$ ($1178 / 1461$). 
This indicates that contamination is concentrated primarily in the new candidates identified solely by photometric clustering. 
The 6.7\% contamination rate of $\mathcal{K}$ shows that even with spectroscopic confirmation, some objects still cannot be excluded as contaminants because their distance measurements and kinematic information are mutually inconsistent.

After removing the unbound stars, two clean sample sets can be defined:
\begin{itemize}
	\item \textbf{Column description of the BHB catalog} $\mathcal{S}_c = \mathcal{C} \setminus \{v_{\rm total} > v_{\rm esc}\}$:
	$|\mathcal{S}_c| = 7776$, retention rate $R_{\rm keep} = 84.2\%$. The complete set of data for this sample is provided in Table ~\ref{tab:catalog_columns}.
	\item \textbf{Contaminant set} $\mathcal{S}_p = \mathcal{C} \cap \{v_{\rm total} > v_{\rm esc}\}$: 
	$|\mathcal{S}_p| = 1461$.
	\item \textbf{Clean spectroscopic BHB sample} $\mathcal{S}_d = \mathcal{K} \setminus \{v_{\rm total} > v_{\rm esc}\}$:
	$|\mathcal{S}_d| = 5072$, retention rate $93.3\%$.
\end{itemize}

  \begin{deluxetable*}{lll}
    \tablecaption{Properties of BHB stars identified in this work. \label{tab:catalog_columns}}
    \tablehead{\colhead{Column} & \colhead{Units} & \colhead{Description}}
    \startdata
\texttt{TARGETID} & -- & DESI MWS DR1 unique target identifier \\
\texttt{RA}       & deg & Right Ascension (J2000) \\
\texttt{DEC}      & deg & Declination (J2000) \\
\texttt{L}        & deg & Galactic longitude \\
\texttt{B}        & deg & Galactic latitude \\
\texttt{MODEL\_U\_CORR} & mag & trimmed SkyMapper $u$-band AB magnitude (synthetic from DESI spectrum) \\
\texttt{MODEL\_V\_CORR} & mag & SkyMapper $v$-band AB magnitude (synthetic) \\
\texttt{MODEL\_G\_CORR} & mag & SkyMapper $g$-band AB magnitude (synthetic) \\
\texttt{MODEL\_R\_CORR} & mag & SkyMapper $r$-band AB magnitude (synthetic) \\
\texttt{MODEL\_I\_CORR} & mag & SkyMapper $i$-band AB magnitude (synthetic) \\
\texttt{MODEL\_Z\_CORR} & mag & SkyMapper $z$-band AB magnitude (synthetic) \\
\texttt{EBV}            & mag & $E(B-V)$ reddening from \citet{1998ApJ...500..525S} \\
\texttt{RV}       & \kms & Barycentric radial velocity (DESI ZBEST \texttt{VRAD}) \\
\texttt{FEH}      & dex  & [Fe/H] metallicity \\
\texttt{FEH\_ERR} & dex  & [Fe/H] uncertainty \\
\texttt{SN\_B}    & --   & DESI B-channel median SNR per pixel ($\sim3600$--5800\,\AA) \\
\texttt{SN\_R}    & --   & DESI R-channel median SNR per pixel ($\sim5760$--7620\,\AA) \\
\texttt{SN\_Z}    & --   & DESI Z-channel median SNR per pixel ($\sim7520$--9824\,\AA) \\
\texttt{PMRA}  & mas\,yr$^{-1}$ & Proper motion $\mu_\alpha\cos\delta$ (Gaia DR3) \\
\texttt{PMDEC} & mas\,yr$^{-1}$ & Proper motion $\mu_\delta$ (Gaia DR3) \\
\texttt{DIST\_BM} & kpc & Heliocentric distance from BHB $M_g$ calibration \\
                  &     & $M_g = -1.43\,(g-i)_0 + 0.23\,\mathrm{[Fe/H]} + 0.49$ \citep{2025ApJS..278...62H} \\
\texttt{R\_GAL}   & kpc & Galactocentric cylindrical radius \\
\texttt{Z\_GAL}   & kpc & Galactocentric height above the plane \\
\texttt{PHI\_GAL} & rad & Galactocentric azimuthal angle, $\phi = \arctan2(y, x)$ \\
\texttt{V\_TOTAL} & \kms & Total Galactocentric space velocity, $\sqrt{V_R^2 + V_\phi^2 + V_Z^2}$ \\
\texttt{V\_ESC}   & \kms & Local escape velocity from \citet{2017MNRAS.465...76M} via \texttt{galpy} \\
\texttt{ETA}      & --   & Rotation-to-pressure-support ratio, $V_\phi\,/\,\sqrt{V_R^2 + V_Z^2}$ \\
\texttt{ECC}     & --  & Orbital eccentricity \\
\texttt{R\_APO}  & kpc & Apocenter distance \\
\texttt{R\_PERI} & kpc & Pericenter distance \\
\texttt{IS\_GSE} & --  & GSE candidate flag: 1 if $e > 0.7$ \citep{2020ApJ...901...48N, 2024MNRAS.527.2165C} \\
\texttt{GC\_CLUSTER}     & -- & SDSS globular cluster name (matched within $3''$; cluster core $r < r_t/3$) \\
\texttt{GC\_IN\_BHB\_BOX} & -- & Flag: 1 if $-0.3 \le g-r \le 0.0$ within the cluster HB magnitude range \\
    \enddata
    \tablecomments{This table is available in its entirety in machine-readable format.}
  \end{deluxetable*}

The intersection $\mathcal{S}_c \cap \mathcal{S}_d$ contains $4685$ stars, accounting for 60.3\% of $\mathcal{S}_c$ and 92.4\% of $\mathcal{S}_d$.
The following analysis focuses on the clean BHB samples ($\mathcal{S}_c$, $\mathcal{S}_d$) and the contaminants ($\mathcal{S}_p$).

It is worth noting that, in the preceding section, a KDE-based Bayesian probability model for BHB stars was constructed, and a probability-weighted purity was defined. This metric reflects the average confidence of the KDE model that the members of a given set are BHB stars, and is equivalent to an expected purity estimate weighted by probability. The calibrated probability-weighted purity of $\mathcal{C}$ given by the KDE model is $\mathcal{P}(\mathcal{C}) = 0.89$.
The escape velocity screening removes $1461$ unbound stars from $\mathcal{C}$ ($15.8\%$), yielding a retention rate $R_{\rm keep} = 0.842$. The small difference between these two values indicates that the escape velocity screening is quite effective at removing contaminants. At the same time, it should be noted that a more precise kinematic filtering may still require a more reliable Galactic gravitational model and improved dynamical parameter measurements.

\begin{figure}[htbp]
	\centering
	\includegraphics[width=0.45\textwidth]{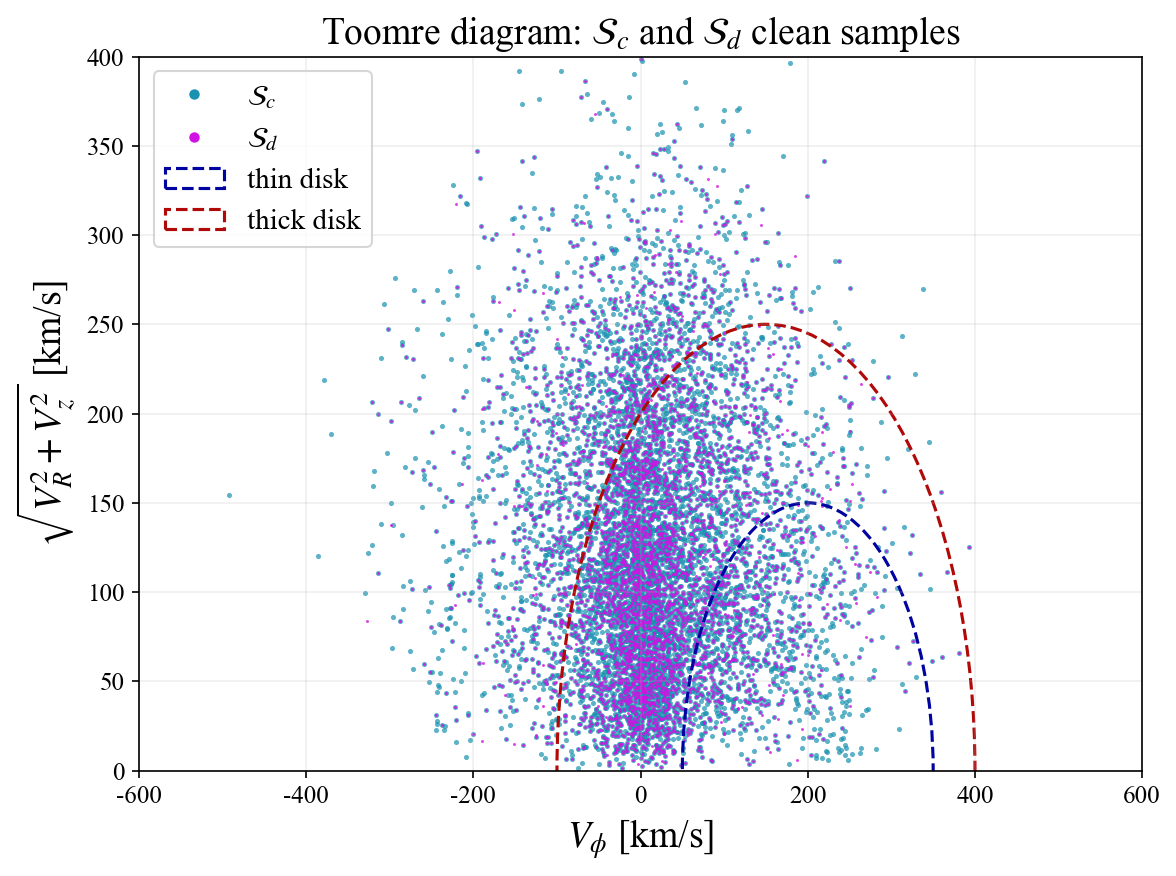}
	\caption{Toomre diagram ($V_\phi$ vs.\ $\sqrt{V_R^2+V_z^2}$) of the clean samples $\mathcal{S}_c$ (blue dots) and $\mathcal{S}_d$ (magenta dots).	The blue and red dashed circles indicate the kinematic boundaries of the thin disk and thick disk, respectively \citep{2019MNRAS.484L..69R}.}
	\label{fig:Toomre_diagram}
\end{figure}

In addition to removing some contaminants through the escape velocity cut, the Toomre diagram, formed by the parameter pair $(V_\phi, \sqrt{V_R^2 + V_z^2})$, serves as an important tool for diagnosing the kinematic membership of stellar populations. 
Figure~\ref{fig:Toomre_diagram} shows the distributions of $\mathcal{S}_c$ (blue dots) and $\mathcal{S}_d$ (magenta dots) in the Toomre plane. 
The bulk of both $\mathcal{S}_c$ and $\mathcal{S}_d$ is concentrated in the typical halo region near $V_\phi \approx 0$, with the combined speed $\sqrt{V_R^2+V_z^2} $ typically spanning $\sim  0$--$300$~km/s. The two distributions overlap strongly, indicating that after the escape velocity screening, the photometric and spectroscopic samples are consistent in their kinematic properties.

With reference to the thin disk (center $(200, 0)$~km/s, radius $\sim$150~km/s) and the thick disk (center $(150, 0)$~km/s, radius $\sim$250~km/s) \citep{2019MNRAS.484L..69R}, the kinematic composition of the 7776 stars in $\mathcal{S}_c$ is: thin disk 1034 stars (13.3\%), thick disk 4,727 stars (60.8\%), and halo 2015 stars (25.9\%).
It should be noted that the thick-disk circle of the criterion adopted by \citet{2019MNRAS.484L..69R} overlaps significantly with the pressure-supported halo population in the region around $V_\phi \approx 0$ and high velocity dispersion; therefore, the apparently high ``thick disk'' fraction is primarily a definitional issue of the criterion and does not represent genuine disk contamination.

Rotation direction statistics reveal that $\mathcal{S}_c$ is dominated by prograde and near-zero rotation: prograde 4416 stars (56.8\%), retrograde 3360 stars (43.2\%), and near-zero rotation ($|V_\phi| < 30$~km/s) 2731 stars (35.1\%). 
The median $V_\phi$ is -10.1~km/s, close to zero rotation, consistent with the expectations of the Gaia-Sausage/Enceladus (GSE) merger event \citep{2018MNRAS.478..611B, 2018Natur.563...85H}. 
This further confirms that the BHB sample obtained through the ``clustering + escape velocity'' selection method has high purity and can serve as an excellent probe dedicated to studying the three-dimensional structure and chemo-dynamical details of the GSE debris.

Quantifying the orbital type with the rotation-to-pressure-support ratio $\eta = V_\phi / \sqrt{V_R^2 + V_z^2}$, the median $|\eta|$ of $\mathcal{S}_c$ is $0.43$, with $\bar{\eta} \approx -0.26 \pm 2.23$. The median $|\eta|$ is far smaller than 1, confirming the dominance of pressure support.

Adopting an eccentricity $e > 0.7$ as the GSE criterion \citep{2020ApJ...901...48N, 2024MNRAS.527.2165C}, we perform orbital integration for all 9,237 stars in $\mathcal{C}$ using the Galactic gravitational potential \citep{2017MNRAS.465...76M} and identify 4,792 GSE candidates, yielding a GSE fraction of 51.9\%. After removing contaminants, the clean subset $\mathcal{S}_c$ has a GSE fraction of 50.3\%. This fraction is consistent with the $\gtrsim 50\%$ inner-halo contribution estimated by \citet{2019MNRAS.486..378L} based on $\sim 3000$ SDSS-Gaia BHB stars. The slight decrease in the GSE fraction after removing contaminants arises because the contaminants are faint objects whose photometric distances are severely overestimated; their amplified tangential velocities make them more prone to acquiring high eccentricities.

It should be noted that the slight dominance of prograde orbits in $\mathcal{S}_c$ (56.8\%) is not in conflict with the high eccentricity characteristic of GSE --- the eccentricity criterion $e>0.7$ constrains only the orbital shape, not the direction of rotation, and GSE debris stars can simultaneously possess high $e$ and either prograde or retrograde $V_\phi$. Hence, the nearly equal fractions of prograde and retrograde orbits and the median $V_\phi$ close to zero are exactly the typical signatures of a well-mixed phase space of GSE merger debris.

Table~\ref{tab:Sp_sigma_profile} presents the evolution of the velocity dispersion components of $\mathcal{S}_c$ with Galactocentric radius $R$. This is one of the most important physical improvements brought about by the escape velocity screening:
in the original $\mathcal{C}$, the velocity dispersion inflates dramatically beyond $R > 30$~kpc (with $\sigma_{V_\phi}$ soaring to $> 4000$~km/s at $R > 100$~kpc), whereas in $\mathcal{S}_c$ the dispersions across all $R$ bins remain at a normal level of $\sim 80$--$110$~km/s, and $\sigma_{V}$ is stable between $90.9$ and 110.8~km/s. At $R > 50$~kpc the dispersions maintain similar values ($\sigma_{V} \approx 92.7$~km/s), but with small number statistics ($N = 76$) and larger uncertainties. Overall, the escape velocity screening completely eliminates the previously observed phenomenon of ``abnormally inflated velocity dispersion at large distances'' in $\mathcal{C}$, confirming that this anomaly is entirely an artifact of distance overestimation contamination rather than a genuine kinematic signal.

\begin{table}[htbp]
	\centering
	\caption{ Velocity dispersion of $\mathcal{S}_c$ as a function of Galactocentric radius}
	\label{tab:Sp_sigma_profile}
	\begin{tabular}{lrrrrr}
		\hline
		$R$ (kpc) & $N$ & $\sigma_{V_\phi}$ & $\sigma_{V_R}$ & $\sigma_{V_z}$ & $\sigma_{V}$ \\
		\hline
		$0$--$10$   & 3005 &101.1 & 113.2 &117.3 &110.8 \\
		$10$--$15$  &1816 & 89.5 &107.2 &94.7 &97.4 \\
		$15$--$20$  &1175 &81.8 &104.7 &84.5 & 90.9 \\
		$20$--$30$  &1177 &95.2 &100.8 &84.3 & 93.7 \\
		$30$--$50$  &523   & 104.9 &100.1 & 83.1  &96.5 \\
		$50$--$100$ & 76    & 105.5 & 84.2  & 87.0  & 92.7 \\
		\hline
	\end{tabular}
	\par\vspace{5pt}
	{\footnotesize \hspace{-6em}\textbf{Note.} $\sigma_{V} \equiv \sqrt{(\sigma_{V_\phi}^2 + \sigma_{V_R}^2 + \sigma_{V_z}^2)\,/\,3}$.}
\end{table}

Table~\ref{tab:Sp_kinematic_summary} summarizes the comparison of core kinematic statistics between $\mathcal{S}_c$ and $\mathcal{C}$.
The median $V_\phi$ of the clean sample (-10.1 vs.\ -11.7~km/s for $\mathcal{C}$) and the median $\sqrt{V_R^2+V_z^2}$ ($116.6$ vs.\ 127.5~km/s) are highly consistent with those of the original sample, indicating that the escape velocity screening does not introduce kinematic bias while removing contaminants --- what is removed are objects with extreme kinematic anomalies, which should not belong to the BHB population in the first place.
The $\sigma$ values of $\mathcal{S}_c$ are significantly narrowed (for instance, the $\sigma$ of $V_\phi$ drops from 1125.2 to 95.3~km/s), reflecting the restoration of physical plausibility.

\begin{table}[htbp]
	\centering
	\caption{Kinematic statistics of $\mathcal{S}_c$ and $\mathcal{C}$}
	\label{tab:Sp_kinematic_summary}
	\begin{tabular}{lrr}
		\hline
		\textbf{Statistic} & \textbf{$\mathcal{C}$} & \textbf{$\mathcal{S}_c$} \\
		\hline
		$N$ & 9237 & 7776 \\
		\hline
		Median $V_\phi$ (km/s) & $-11.7$ & $-10.1$ \\
		Median $\sqrt{V_R^2+V_z^2}$ (km/s) & $127.5$ & $116.6$ \\
		$\sigma_{V_\phi}$ (km/s) & $1125.2$ & $95.3$ \\
		GSE fraction ($e > 0.7$) & 51.9\% & 50.3\% \\
		Prograde fraction & 56.8\% & 56.8\% \\
		\hline
	\end{tabular}
\end{table}

\begin{figure}[htbp]
	\centering
	\includegraphics[width=0.45\textwidth]{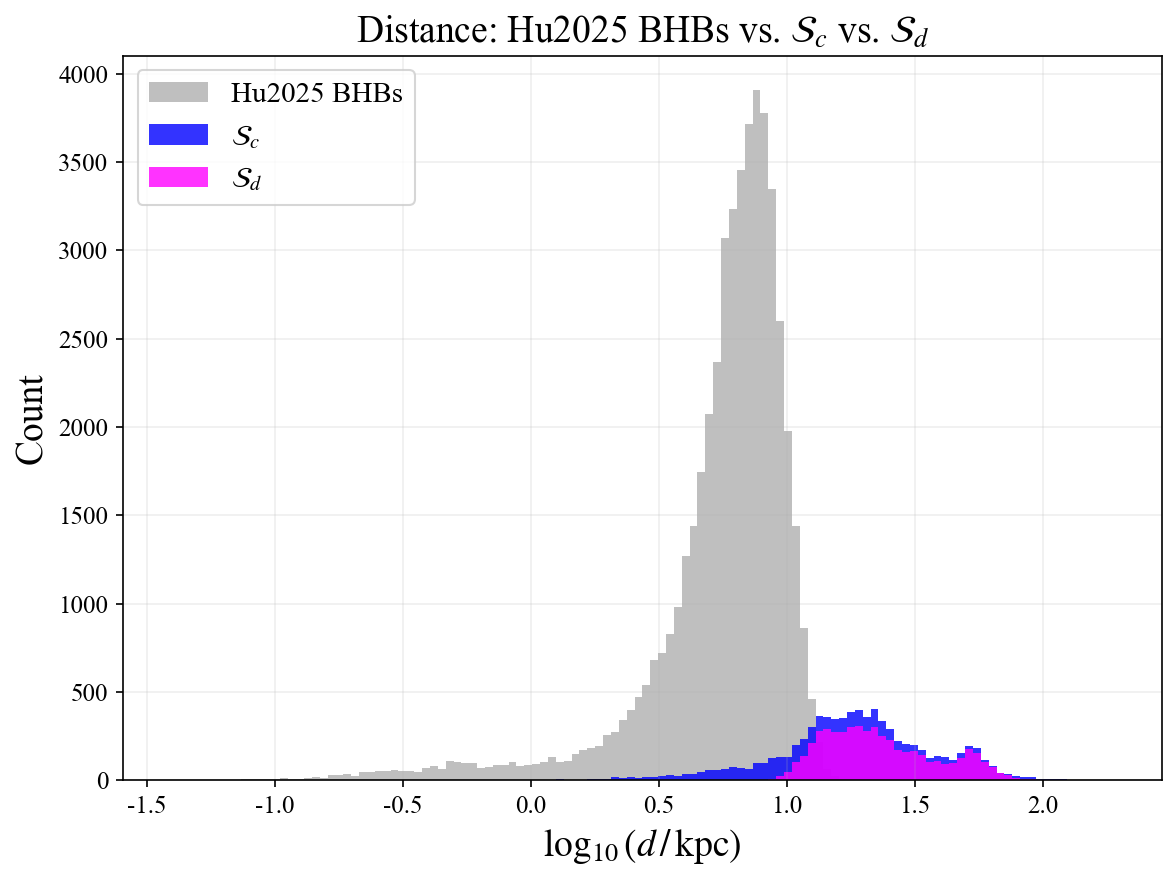}
	\caption{Distance distribution histograms for clean photometric $\mathcal{S}_c$ (blue), clean spectroscopic $\mathcal{S}_d$ (magenta), and Hu2025 BHBs (gray), shown in logarithmic scale.}
	\label{fig:distance_clean}
\end{figure}

Figure~\ref{fig:distance_clean} shows the distance distribution characteristics of $\mathcal{S}_c$ (blue), $\mathcal{S}_d$ (magenta), and the Hu2025 BHBs (gray) samples.
The median distance of $\mathcal{S}_c$ is $18.7$~kpc, extending to $190.6$~kpc;
that of $\mathcal{S}_d$ is $21.3$~kpc, with a maximum of $103.5$~kpc.
Compared with the original $\mathcal{C}$ (median $21.3$~kpc, maximum $664.1$~kpc),
the escape velocity screening removes the $1461$ contaminant stars with severely overestimated distances,
bringing the distance distribution of the sample back to a physically reasonable range.
$\mathcal{S}_c$ still covers a broad radial sequence from the solar neighborhood ($< 5$~kpc) to the outer halo ($> 100$~kpc),
but the number of stars with $d > 100$~kpc drops sharply from about $508$ in $\mathcal{C}$ to $25$,
confirming that distance overestimation at the extreme far end is the main physical source of contamination.

\begin{figure}[htbp]
	\centering
	\includegraphics[width=0.40\textwidth]{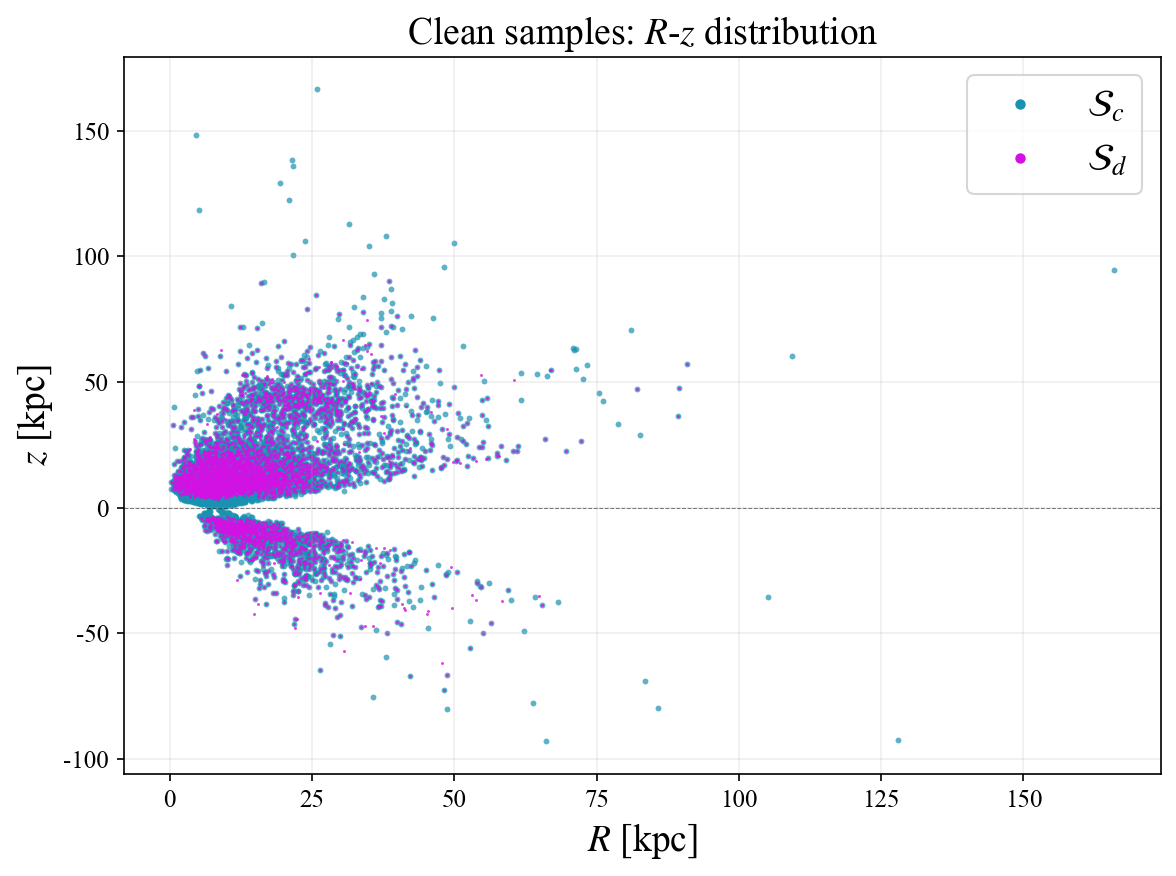}
	\caption{Spatial distribution of the clean samples in the Galactocentric cylindrical coordinate system $(R, z)$.
		$\mathcal{S}_c$ (blue dots) and $\mathcal{S}_d$ (magenta dots).
		The two distributions overlap strongly, with $\mathcal{S}_c$ extending farther in both the radial and vertical directions, demonstrating the advantage of the photometric method in probing depth.}
	\label{fig:Rz_distributions}
\end{figure}

Figure~\ref{fig:Rz_distributions} shows the $(R, z)$ distributions of $\mathcal{S}_c$ and $\mathcal{S}_d$: the blue dots represent $\mathcal{S}_c$, and the magenta dots represent $\mathcal{S}_d$. The spatial distributions of the two are highly consistent: the median $R$ and $|z|$ of $\mathcal{S}_c$ are $12.4$~kpc and $12.1$~kpc, respectively; the median $R$ and $|z|$ of $\mathcal{S}_d$ are $13.9$~kpc and $13.6$~kpc, respectively. $\mathcal{S}_d$ has 53 stars at $R > 50$~kpc, while $\mathcal{S}_c$ has 80 stars ($1.0\%$) in the same region. These distant objects, which remain physically bound to the Galactic gravitational potential after the escape velocity screening, are valuable tracers for probing the outer halo structure.

Compared with the original $\mathcal{C}$, the maximum $R$ of $\mathcal{S}_c$ shrinks from $495.7$~kpc to $165.9$~kpc, and the maximum $|z|$ shrinks from $446.7$~kpc to $166.7$~kpc.
This contraction is not a simple truncation, but rather the precise identification of distance-overestimation contamination by the escape velocity screening --- those contaminant stars whose distances were overestimated to hundreds of kpc also have excessively amplified tangential velocities, and are precisely the dominant population with $v_{\rm total} > v_{\rm esc}$.

It is noteworthy that the distances of $\mathcal{S}_c$ (and $\mathcal{S}_d$) in Figure~\ref{fig:distance_clean} exhibit a bimodal distribution, with one peak at about $18$~kpc and the other at about $50$~kpc, implying the possible presence of two components in the sample. In Figure~\ref{fig:Rz_distributions}, a high-density BHB clump is visible at $z \approx 30$--$55$~kpc and $R \approx 10$--$40$~kpc. This structure coincides remarkably with the northern trailing arm of the Sagittarius stream \citep{2024ApJ...963...95C}, suggesting that the clump is part of this well-known outer halo substructure.

\section{Statistical Characteristics of the Contaminants}
\label{sec:BHB_contam_stat}

\begin{figure*}[htbp]
	\centering
	\includegraphics[width=0.95\textwidth]{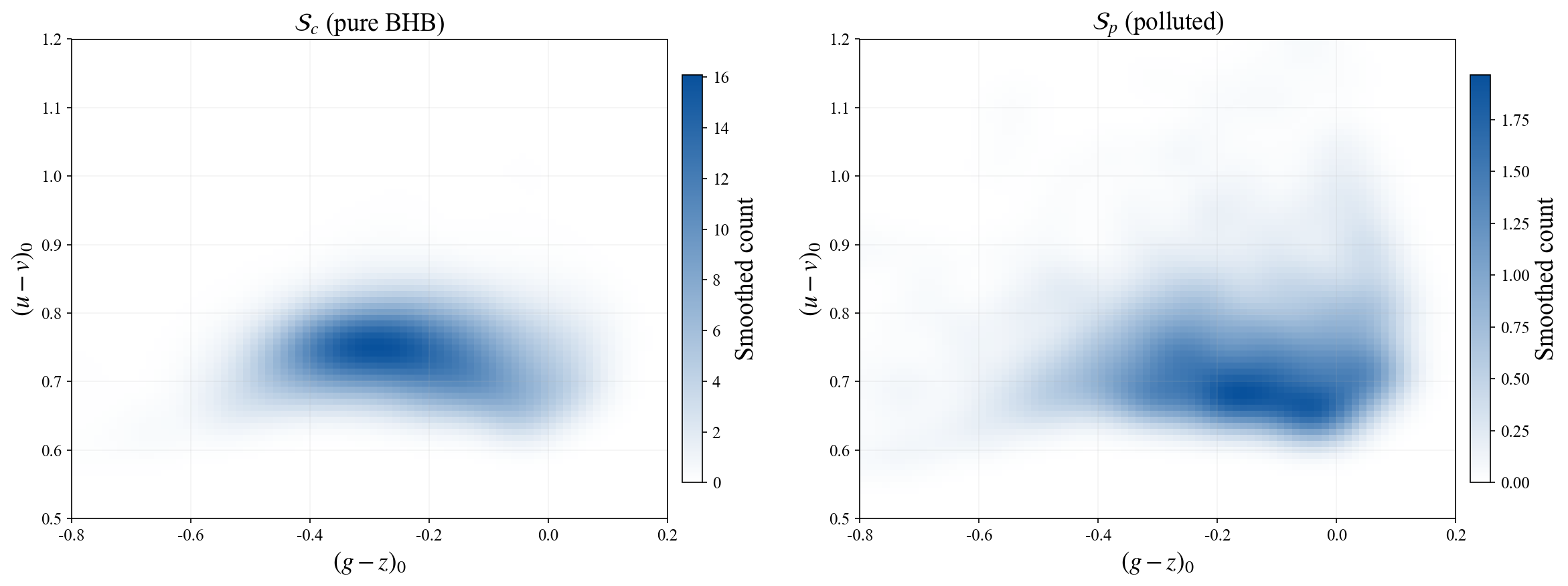}
	\caption{Independently normalized smoothed density distributions of $\mathcal{S}_c$ (left, $|\mathcal{S}_c|=7776$) and $\mathcal{S}_p$ (right, $|\mathcal{S}_p|=1461$) in the $(g-z)_0$--$(u-v)_0$ color space. The contaminants display a strongly asymmetric density profile. While the genuine BHB density contours remain compact and centrally peaked, the contaminant contours are elongated and diluted, with the highest relative density shifting toward the cluster's boundary in the direction of $(g-z)_0$ reddening and $(u-v)_0$ dispersion.}
	\label{fig:density_compare}
\end{figure*}

We systematically diagnose the contaminants from the perspective of color-space distribution by comparing the independently normalized smoothed density distributions of $\mathcal{S}_c$ and $\mathcal{S}_p$ in the $(g-z)_0$--$(u-v)_0$ color space (Figure~\ref{fig:density_compare}), in order to determine whether the contaminants are distinguishable in photometric colors and to trace their physical origin. The left panel of Figure~\ref{fig:density_compare} shows the density distribution of the clean BHB sample after removing contaminants, while the right panel shows that of the contaminants. As can be seen from the comparison, the contaminants exhibit a markedly skewed and diffuse density distribution that is not centrally concentrated. Unlike the symmetric BHB profile, the contaminant density builds toward the redder, low-gravity periphery of the BHB locus, forming an extended wing that leaks into the non-BHB region. This morphology strongly suggests that these contaminants represent a natural dispersion from the background population into the BHB selection boundary, rather than a physically distinct cluster.

Table~\ref{tab:color_stats} summarizes the color-space statistics for the four sets. The differences in median $(g-z)_0$ and $(u-v)_0$ between $\mathcal{S}_c$ and $\mathcal{K}$ are only $+0.008$~mag and $-0.003$~mag, respectively, and the two distributions are highly consistent (KS $D < 0.05$), indicating that the $v_{\rm esc}$ screening does not introduce systematic bias.

\begin{table}[htbp]
	\centering
	\caption{Summary of color-space statistics}
	\label{tab:color_stats}
	\begin{tabular}{l c c c c}
		\hline
		\multirow{2}*{\textbf{Sample}} & \multicolumn{2}{c}{$\mathbf{(g-z)_0}$} & \multicolumn{2}{c}{$\mathbf{(u-v)_0}$} \\
		\cline{2-3} \cline{4-5}
		& Median & $\sigma_{gz}$ & Median & $\sigma_{uv}$ \\
		\hline
		$\mathcal{K}$     & $-0.246$ & $0.144$ & $0.739$ & $0.063$ \\
		$\mathcal{C}$     & $-0.229$ & $0.158$ & $0.735$ & $0.065$ \\
		$\mathcal{S}_c$   & $-0.238$ & $0.152$ & $0.736$ & $0.053$ \\
		$\mathcal{S}_p$   & $-0.165$ & $0.184$ & $0.725$ & $0.106$ \\
		\hline
	\end{tabular}
\end{table}

The statistical information shows that the median color offsets of $\mathcal{S}_p$ relative to $\mathcal{S}_c$ are small: $(g-z)_0$ exhibits a median redward shift of $\Delta = +0.073$~mag, and $(u-v)_0$ a blueward shift of $\Delta = -0.011$~mag. However, the color dispersion of $\mathcal{S}_p$ is significantly larger: $\sigma_{\rm uv} = 0.106$ for $(u-v)_0$, which is $2.0$ times that of $\mathcal{S}_c$ ($0.053$), and $\sigma_{\rm gz} = 0.184$ for $(g-z)_0$, which is $1.2$ times that of $\mathcal{S}_c$ ($0.152$). This means that most contaminants are indistinguishable from BHB stars in color and cannot be removed by simple color cuts.

\begin{figure*}[htbp]
	\centering
	\includegraphics[width=0.95\textwidth]{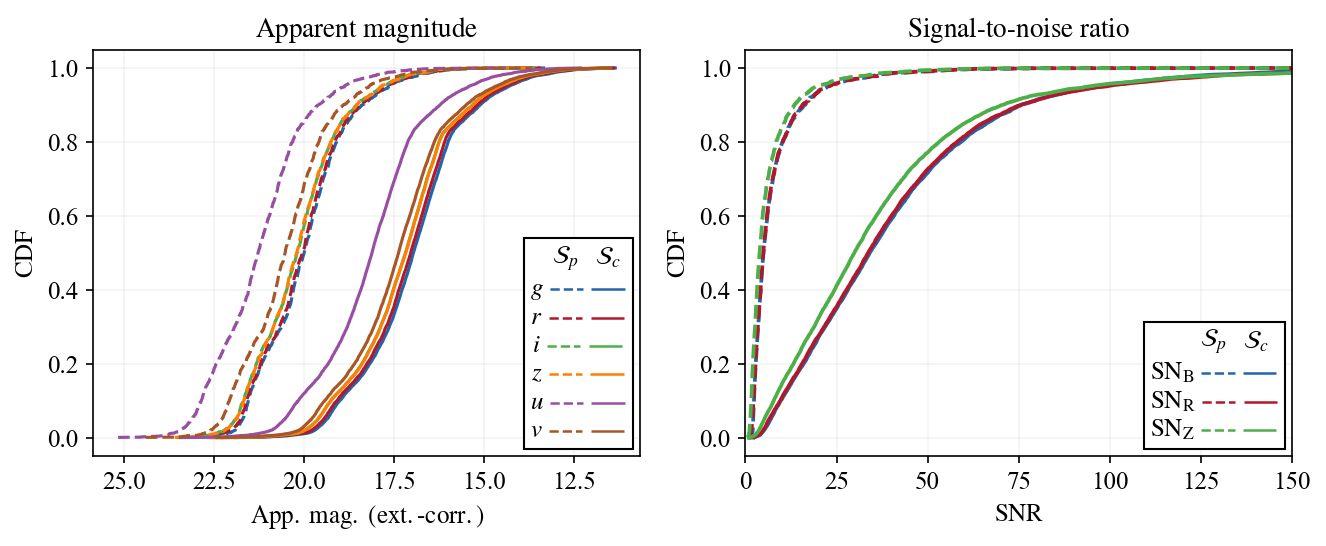}
	\caption{Cumulative distribution comparison of apparent magnitudes (left) and signal-to-noise ratios (right) between the clean sample $\mathcal{S}_c$ (solid lines) and the contaminants $\mathcal{S}_p$ (dashed lines).
		Left: extinction-corrected apparent magnitudes in the $g$, $r$, $i$, $z$, $u$, and $v$ bands; the magnitude axis is inverted so that brighter sources appear toward the left.
		Right: signal-to-noise ratios in the B (blue), R (red), and Z (green) channels of DESI.
		The contaminants are systematically fainter than the clean sample by $\sim 3$~mag across all bands, and their signal-to-noise ratios are consistently concentrated at very low values in all three channels.}
	\label{fig:snr_contamination}
\end{figure*}

Furthermore, the distances of the contaminant stars removed by the escape velocity screening $\mathcal{S}_p$ are generally overestimated. Combined with the standard-candle nature of BHB stars, this implies that these contaminants should have relatively faint observed brightnesses and low signal-to-noise ratios. This can be easily verified from the data.

We conduct a systematic comparison of the apparent magnitudes and spectroscopic signal-to-noise ratios between the contaminant stars rejected by the escape velocity screening $\mathcal{S}_p$ and the clean photometric BHB sample $\mathcal{S}_c$. Figure~\ref{fig:snr_contamination} shows, in the left and right panels respectively, the cumulative distribution functions of the extinction-corrected apparent magnitudes in the six bands ($grizuv$, left panel) and the signal-to-noise ratios in the three DESI channels (SN\textsubscript{B}, SN\textsubscript{R}, SN\textsubscript{Z}, right panel). $\mathcal{S}_c$ is represented by solid lines, and $\mathcal{S}_p$ by dashed lines, with the same color used for the same band.

The apparent magnitude distribution reveals a key feature: the median magnitudes of $\mathcal{S}_p$ are systematically fainter than those of $\mathcal{S}_c$ by about $3.0$--$3.2$~mag in all six bands (for example, in the $g$ band: $20.00$ vs.\ $16.97$). The six CDF curves of the contaminants are clearly separated from the six CDF curves of the clean sample, indicating that the contaminants are not the result of random errors in the BHB standard-candle distance estimation, but rather a population of systematically fainter objects --- when the BHB absolute magnitude calibration is incorrectly applied to these intrinsically fainter objects, their photometric distances are systematically overestimated.

The signal-to-noise ratio distributions show a consistent and even more extreme separation: the median SNRs of $\mathcal{S}_p$ in all three bands are below 5 (SN\textsubscript{B}=5.0, SN\textsubscript{R}=4.9, SN\textsubscript{Z}=3.9), while those of $\mathcal{S}_c$ are 34.1, 33.6, and 29.9, respectively; the KS test yields $D = 0.70$--$0.71$ across the three bands ($p \ll 0.001$). The high degree of consistency across the three bands indicates that the low SNR is a common spectral quality issue affecting the entire wavelength range.

Summarizing the above analyses, the physical picture of the contamination is self-consistent: the contaminants are a class of faint objects near the detection limit of DESI MWS, whose low spectroscopic signal-to-noise ratios directly result from their fainter apparent magnitudes. The low photon counts amplify the photometric errors, manifested as a $(u-v)_0$ color dispersion about twice that of the clean BHB sample. When these faint stars happen to fall within the BHB region in the $(g-z)_0$--$(u-v)_0$ color space, the clustering algorithm identifies them as BHB candidates. The BHB absolute magnitude calibration is then incorrectly applied to these intrinsically fainter objects, inevitably leading to overestimated distances for a subset of them, and consequently their space velocities are amplified beyond the escape velocity. Thus, the $v > v_{\rm esc}$ screening essentially targets the physical mechanism of ``photometric distance overestimation caused by faint contaminants,'' rather than a simple rejection of kinematic outliers. It should be emphasized that this screening mechanism depends on the accuracy of the Galactic gravitational model; the  \citet{2017MNRAS.465...76M} potential adopted here is relatively conservative, with comparatively high escape velocities especially at large distances.

\section{External Purity Validation Based on Globular Cluster H-R Diagrams}
\label{sec:sdss_cluster_test}

The probabilistic model validation (Section~\ref{sec:prob_model_eval}) and the escape velocity screening (Section~\ref{sec:de_contamination}) constrain the sample purity from a purely probabilistic perspective and a dynamical perspective, respectively; however, they rely on the completeness of the training data and the correctness of the Galactic gravitational potential model, respectively. 
SDSS globular cluster H-R diagrams provide a completely independent means of verification: BHB members of globular clusters can be reliably identified by their positions on the $(g-r,\ g)$ H-R diagram \citep{2012AJ....143...86V}, because stars within a cluster share a common distance and metallicity, making the morphology of the horizontal branch clearly distinguishable.

We perform positional cross-matching of $\mathcal{S}_c$ (clean photometric BHB) and $\mathcal{S}_d$ (clean spectroscopic BHB) with the point-source photometry in the fields of 12 SDSS globular clusters \citep{2008ApJS..179..326A}. Among them, 6 clusters with good SDSS photometric coverage (NGC~5024/M53, NGC~5053, NGC~5466, NGC~2419, Pal~5, and NGC~6341/M92) yield BHB matches. We first extract all photometric point sources in the core region (within a radius of $1/3$ of the tidal radius $r_t$) of each cluster from the SDSS catalog and plot the $(g-r,\ g)$ H-R diagram as the background; we then match the total MWS DR1 sample with SDSS photometric stars using a $3''$ radius and mark their positions on the diagram. Finally, we check whether these matched stars satisfy $-0.3 \le g-r \le 0.0$ and have $g$-band magnitudes within the typical range of the cluster horizontal branch \citep{2012AJ....143...86V}.

	\begin{figure*}[htbp]
			\centering
			\begin{minipage}[c]{0.95\textwidth}
					\centering
					\includegraphics[width=\textwidth]{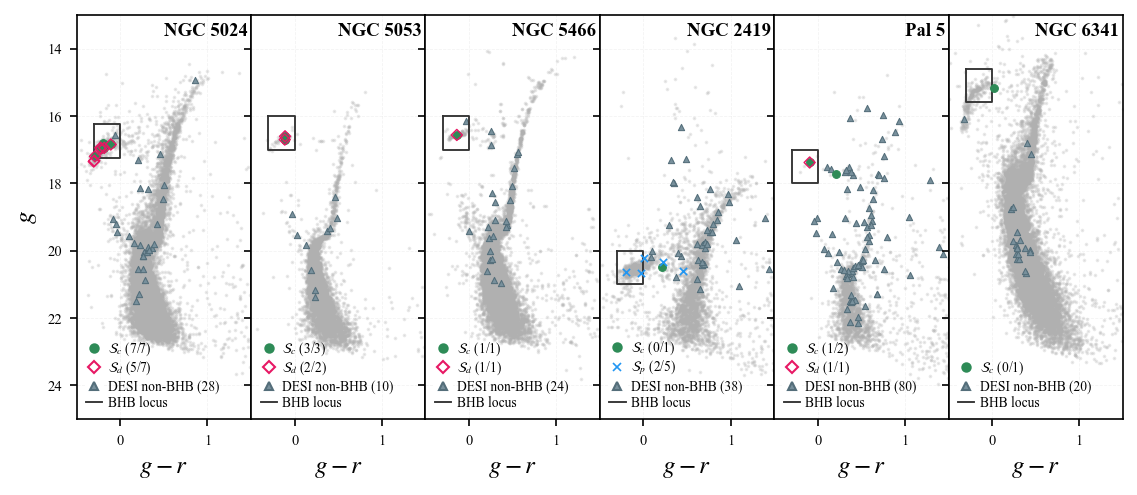}
				\end{minipage}
			\caption{Independent purity validation based on SDSS globular cluster H-R diagrams. The 6 panels correspond to 6 globular clusters with good SDSS photometric coverage.
				Gray dots are all SDSS photometric point sources in the core region ($r < r_t/3$) of each cluster, forming the background stellar distribution on the $(g-r,\ g)$ H-R diagram. 
				All DESI MWS DR1 sources were cross-matched with these SDSS fields using a $3''$ matching radius; the matched points are overlaid: green filled circles mark sources from the clean photometric BHB sample $\mathcal{S}_c$ that match the cluster field within $3''$, and magenta open diamonds mark the matched sources from the clean spectroscopic BHB sample $\mathcal{S}_{\rm d}$. Blue crosses indicate matched noise (contaminant) sources, and triangles denote all other matched DESI non-BHB stars. The black boxes outline the typical BHB region ($-0.3 \le g-r \le 0.0$, with $g$-band magnitude within the cluster horizontal branch range).}
			\label{fig:cluster_hr_diagram}
	\end{figure*}

\begin{table*}[htbp]
	\centering
	\caption{SDSS globular cluster BHB cross-match statistics}
	\label{tab:sec6_purity}
	\begin{tabular}{lcccc}
		\hline
		\multirow{2}{*}{Cluster} &
		\multicolumn{2}{c}{Photometric sample} &
		\multicolumn{2}{c}{Spectroscopic sample} \\
		\cline{2-3} \cline{4-5}
		& $\mathcal{C}$ & $\mathcal{S}_c$ & $\mathcal{K}$ & $\mathcal{S}_{\rm d}$ \\
		\hline
		NGC~5024 (M53) & 7/7 (100\%) & 7/7 (100\%) & 5/7 (71\%) & 5/7 (71\%) \\
		NGC~5053       & 3/3 (100\%) & 3/3 (100\%) & 2/2 (100\%) & 2/2 (100\%) \\
		NGC~5466       & 1/1 (100\%) & 1/1 (100\%) & 1/1 (100\%) & 1/1 (100\%) \\
		Pal~5          & 1/2 (50\%)  & 1/2 (50\%)  & 1/1 (100\%) & 1/1 (100\%) \\
		NGC~2419       & 2/6 (33\%)  & 0/1 (0\%)   & 0/1 (0\%)   & -- \\
		NGC~6341 (M92) & 0/1 (0\%)   & 0/1 (0\%)   & --          & -- \\
		\hline
		\textbf{Total} & \textbf{14/20 (70.0\%)} & \textbf{12/15 (80.0\%)} &
		\textbf{9/12 (75.0\%)} & \textbf{9/11 (81.8\%)} \\
		\hline
	\end{tabular}
\end{table*}

Figure~\ref{fig:cluster_hr_diagram} displays the $(g-r,\ g)$ H-R diagrams for the 6 clusters, i.e., the BHB matching results. The gray dots represent SDSS photometric stars in the core region, the green filled circles indicate BHB stars matched with $\mathcal{S}_c$, the magenta open diamonds indicate stars matched with $\mathcal{S}_{\rm d}$, and the black boxes outline the theoretical BHB region. In addition, the contaminant sample $\mathcal{S}_p$ is marked with blue crosses, while all matched non-BHB data from DESI DR1 are marked with blue triangles. The matched BHB sample statistics for each cluster are summarized in Table~\ref{tab:sec6_purity}. $\mathcal{S}_c$ has a total of 15 matched sources, of which 12 lie within the black boxes, and the remaining 3 outside the boxes are verified by SIMBAD as RR Lyrae variables. $\mathcal{S}_{\rm d}$ has a total of 11 matched sources, of which 9 lie within the black boxes, and the other 2 outside the boxes but very close to the box boundary cannot be verified via SIMBAD. Another 5 contaminant stars removed by the escape velocity cut are successfully matched, all of which are in the direction of NGC~2419 and are likely foreground stars of this cluster. In addition, NGC 5024 and NGC 5466 each contain one non-BHB source located inside the respective box. SIMBAD coordinate queries confirm that only the source inside the box of NGC 5466 is matched to an RR Lyrae variable. Although the match numbers are relatively well for NGC~5024, the numbers for the other clusters are too low to yield high statistical significance. Nevertheless, overall, the purity of $\mathcal{C}$ is slightly lower than that of $\mathcal{K}$, while after removing contaminants, the purity of $\mathcal{S}_c$ is comparable to that of $\mathcal{S}_{\rm d}$. This qualitative conclusion is consistent with the design expectation of the escape velocity screening---to remove kinematically anomalous stars rather than color-space outliers.

\section{Summary and Discussion}
\label{sec:discussion}

Based on the SkyMapper $(u,v,g,z)$ photometric colors synthesized from DESI MWS DR1 spectroscopy, together with proper motions and radial velocities, this paper constructs a complete methodological framework for the selection and quality control of a photometric BHB catalog. First, using HDBSCAN density-based clustering in the $(g-z)_0$--$(u-v)_0$ color space, we identify 9,237 BHB candidates ($\mathcal{C}$) from 61,374 bright stars. We then introduce a capture-recapture method and a KDE Bayesian probability model to assign a calibrated posterior probability to each star, enabling quantitative evaluation of purity and completeness within a ``soft classification'' framework. 
Subsequently, using a Galactic gravitational potential model, we compute the local escape velocity $v_{\rm esc}(R,z)$ and remove 1,461 contaminant stars ($\mathcal{S}_p$) whose space velocities exceed the escape velocity, constructing the clean photometric BHB sample $\mathcal{S}_c$ ($|\mathcal{S}_c|=7776$). 
We further conduct a systematic investigation of the physical nature of the contaminants through two complementary approaches: signal-to-noise ratio analysis and comparison of color-space distributions. 
The analysis reveals that the contaminants filtered out by the escape-velocity cut are predominantly a class of faint objects at the detection limit of DESI MWS.

The core methodological strategy of this paper---``color-space clustering + 6D phase-space escape velocity screening''---achieves, by design, a complementarity between completeness and physical self-consistency.  HDBSCAN clustering greatly extends the detection depth in the low signal-to-noise regime, but this comes at the cost of mixing in a large number of contaminants: in $\mathcal{S}_3$ (the clustering-only subset, $|\mathcal{S}_3|=4269$), the fraction of stars with $v > v_{\rm esc}$ reaches as high as 27.6\%, contributing 80.6\% of unbound stars in $\mathcal{C}$.  The escape velocity screening serves precisely as a physical filter targeting this shortcoming---it does not rely on external spectroscopic confirmation, but instead uses only the dynamical constraint of the Galactic gravitational potential to flag the 1,461 unbound stars in $\mathcal{C}$ as physically implausible. After screening, $\mathcal{S}_c$ maintains a stable three-dimensional velocity dispersion $\sigma_{V}$ in the range $90.9$--$110.8$~km/s across all Galactocentric radius bins, completely eliminating the artificial inflation of $\sigma_{V} > 500$~km/s seen in $\mathcal{C}$ at $R > 30$~kpc.  At the same time, the escape velocity screening does not alter the core kinematic properties of the sample---the median $V_\phi$ ($-10.1$ vs.\ $-11.7$~km/s) and the median $\sqrt{V_R^2+V_z^2}$ ($116.6$ vs.\ $127.5$~km/s) of $\mathcal{S}_c$ and $\mathcal{C}$ are highly consistent, and the prograde fraction remains 56.8\% before and after screening (Table~\ref{tab:Sp_kinematic_summary}), indicating that the screening removes extreme outliers distributed in the wings of the velocity space rather than any specific kinematic component, and therefore introduces no selection bias.

The KDE probability model provides an independent cross-validation of the above strategy from the color-space perspective. 
The probability-weighted purity of $\mathcal{C}$ is 0.89, suggesting that approximately 89\% of the clustering-identified candidates are genuine BHB stars. 
This is broadly consistent with the escape-velocity retention rate $R_{\rm keep}=0.842$, which indicates that about 84\% of the candidates are dynamically bound to the Galaxy. 
Although the two metrics have different statistical meanings---purity measures the fraction of true BHB stars, while retention rate measures the fraction surviving the kinematic cut---their rough agreement shows that the two completely independent approaches, color-based probabilistic classification and dynamics-based physical screening, converge on a comparable assessment of the overall sample reliability. 

Furthermore, the comparison of signal-to-noise ratios and color-space distributions yields a consistent physical picture of the contaminants. 
On the one hand, the median SNRs of $\mathcal{S}_p$ in all three DESI channels are below 5 (SN\textsubscript{B}=5.0, SN\textsubscript{R}=4.9, SN\textsubscript{Z}=3.9), showing an extreme separation from $\mathcal{S}_c$ (SN\textsubscript{B}=34.1) with KS distances $D=0.70$--$0.71$. Correspondingly, the contaminants are systematically fainter than $\mathcal{S}_c$ by about $3.0$--$3.2$~mag in all six bands. This gap places the contaminants firmly at the detection limit of DESI MWS. 
On the other hand, in the $(g-z)_0$--$(u-v)_0$ color space, although the contaminants exhibit a more dispersed distribution than $\mathcal{S}_c$ (the $\sigma$ of $(u-v)_0$ is twice as large), the median color offset is merely $\Delta(g-z)_0=+0.073$~mag. 
Together, these two facts---extremely low SNR and near-indistinguishable colors---explain both why the contaminants were initially selected by the clustering algorithm (their colors mimic those of genuine BHB stars) and why they were ultimately identified as unphysical (low SNR leads to overestimated photometric distances and inflated space velocities).

These pieces of evidence point to a unified physical mechanism: for faint objects, low photon counts amplify the random errors in the synthetic photometric colors; although their mean colors may still fall within the BHB region and thus be selected by the clustering algorithm, the BHB absolute magnitude calibration is erroneously applied to these intrinsically fainter objects, causing their photometric distances to be systematically overestimated and their space velocities consequently amplified beyond the escape velocity. 
Therefore, the escape velocity screening is essentially a physical filter targeting ``low-SNR--driven photometric distance overestimation,'' rather than a simple removal of kinematic outliers. 
It should be noted that the effectiveness of this screening depends on the accuracy of the Galactic potential model---the McMillan (2017) model adopted in this paper yields relatively conservative escape velocities at large distances.

As an independent purity check, we analyzed the $(g-r, g)$ H-R diagrams of six SDSS globular clusters. 
The aggregate purity of $\mathcal{S}_c$ is 80.0\% (12/15), comparable to the 81.8\% (9/11) of the clean spectroscopic sample $\mathcal{S}_d$. 
Although the statistics from the globular cluster cross-match are limited, this result is broadly consistent with the purity assessments from the color-based probability model and the escape velocity screening presented earlier.

In terms of sample size and quality, compared with previous BHB samples based on SDSS/SEGUE spectroscopy \citep{2008ApJ...684.1143X,2011ApJ...738...79X}, the 7,776 BHB stars constructed in this work reach a greater spatial depth. 
Compared with the $\sim$5,500 BHB stars spectroscopically confirmed by \citet{2025MNRAS.542..560B} from DESI DR1, $\mathcal{S}_c$ expands the sample size by about 43\% while maintaining comparable purity. 
At the methodological level, compared with the purely photometric selection of \cite{2012AJ....143...86V} based on SDSS $u-g$ color cuts (completeness 50\%--70\%, contamination rate $\sim$30\%), the three-step strategy of this work---density-based clustering + probability weighting + escape velocity screening---achieves a significant improvement in both purity and completeness.
This high-purity, wide-coverage BHB catalog lays the data foundation for future chemo-dynamical studies of the Galactic halo and the detailed mapping of the Gaia-Sausage/Enceladus merger debris.

\section*{Acknowledgments}

This work was supported by the National Natural Science Foundation of China under grant No. 12573026, the National Key R$\rm\&$D Program of China 2024YFA1611903, and the China Manned Space Project under grant No. CMS-CSST-2025-A13. 
We also acknowledge that this research is funded by Science Research Project of Hebei Education Department (BJ2026091, CYZD2026002). 

The authors acknowledge the use of DeepSeek for literature search, language polishing, and programming assistance during the preparation of this work. All generated content was reviewed, edited, and fully vetted by the authors, who assume full responsibility for the final manuscript.

This research used data obtained with the Dark Energy Spectroscopic Instrument (DESI). DESI construction and operations is managed by the Lawrence Berkeley National Laboratory. This material is based upon work supported by the U.S. Department of Energy, Office of Science, Office of High-Energy Physics, under Contract No. DE–AC02–05CH11231, and by the National Energy Research Scientific Computing Center, a DOE Office of Science User Facility under the same contract. Additional support for DESI was provided by the U.S. National Science Foundation (NSF), Division of Astronomical Sciences under Contract No. AST-0950945 to the NSF’s National Optical-Infrared Astronomy Research Laboratory; the Science and Technology Facilities Council of the United Kingdom; the Gordon and Betty Moore Foundation; the Heising-Simons Foundation; the French Alternative Energies and Atomic Energy Commission (CEA); the National Council of Humanities, Science and Technology of Mexico (CONAHCYT); the Ministry of Science and Innovation of Spain (MICINN), and by the DESI Member Institutions: www.desi.lbl.gov/collaborating-institutions. The DESI collaboration is honored to be permitted to conduct scientific research on I’oligam Du’ag (Kitt Peak), a mountain with particular significance to the Tohono O’odham Nation. Any opinions, findings, and conclusions or recommendations expressed in this material are those of the author(s) and do not necessarily reflect the views of the U.S. National Science Foundation, the U.S. Department of Energy, or any of the listed funding agencies.

	\bibliographystyle{aasjournal}
	\bibliography{ref_list}

\end{document}